\documentclass[12pt]{article}
\usepackage[textwidth=6.25in]{geometry}

\usepackage{enumerate}
\usepackage{amsmath} 
\usepackage{times}
\usepackage{amsmath,amsthm,amssymb}
\usepackage{fancyhdr}
\usepackage{moreverb}
\usepackage{graphicx}
\usepackage{amssymb}
\usepackage{url}
\usepackage{multirow} 
\usepackage[boxed]{algorithm}
\usepackage{algpseudocode}
\usepackage{multirow} 
\usepackage{geometry}
\usepackage{fix-cm}
\usepackage{natbib}
\usepackage{caption}
\usepackage{subcaption}
\usepackage{color}
\usepackage{bbold}
\usepackage{setspace}
\usepackage{titlesec}
\titleformat{\section}[hang]{\centering \bfseries\large}{\thesection.}{0.4em}{}

\newcommand{\R}{\mathbb{R}}
\newcommand{\PP}{{\bf P}}

\renewcommand{\P}{\mathbb{P}}
\newcommand{\E}{\mathbb{E}}

\newtheorem{theorem}{Theorem}[section]

\newtheorem{proposition}[theorem]{Proposition}

\DeclareMathOperator*{\argmax}{arg\,max}
\DeclareMathOperator*{\argmin}{arg\,min}

\theoremstyle{definition}

\theoremstyle{definition}

\def\one{{\bf 1}}

\def\bSigma{{\bf \Sigma}}

\def\bOmega{{\bf \Omega}}

\def\dd{{\bf d}}
\def\D{{\bf D}}
\def\v{{\bf v}}

\def\m{{\bf m}}

\def\e{{\bf e}}
\def\x{{\bf x}}

\def\A{{\bf A}}
\def\M{{\bf M}}
\def\X{{\bf X}}
\def\Q{{\bf Q}}

\def\Z{{\bf Z}}

\def\SS{{\bf S}}
\def\I{{\bf I}}
\def\F{{\cal F}}

\def\P{{\mathbb P}}
\def\Qp{{\mathbb Q}}
\def\E{{\mathbb E}}
\def\Var{{\rm Var}\,}
\def\Cov{{\rm Cov}\,}

\def\|{\, | \,}

\def\cond{\text{cond}}
\def\diag{\text{diag}}

\def\vec{\text{vec}}
\def\Diag{\text{Diag}}
\def\diag{\text{diag}}
\def\Tr{\text{tr}}
\def\supp{\text{supp}}

\def\logit{{\rm logit}}

\title{\singlespace
 \vspace{-3em} \Large Partial Information Framework: Model-Based Aggregation of Estimates from Diverse Information Sources}
\author{\vspace{5em}
Ville A. Satop\"a\"a, Shane T. Jensen, Robin Pemantle, and Lyle H. Ungar \thanks{Ville A. Satop\"a\"a is a Doctoral Candidate, Department of Statistics, The Wharton School of the University of Pennsylvania, Philadelphia, PA 19104-6340 (e-mail: satopaa@wharton.upenn.edu); Shane T. Jensen is a Statistician, Department of Statistics, The Wharton School of the University of Pennsylvania, Philadelphia, PA 19104-6340 (e-mail: stjensen@wharton.upenn.edu); Robin Pemantle is a Mathematician, Department of Mathematics, University of Pennsylvania, Philadelphia, PA 19104-6395 (e-mail: pemantle@math.upenn.edu); Lyle H. Ungar is a Computer Scientist, Department of Computer and Information Science, University of Pennsylvania, Philadelphia, PA 19104-6309 (e-mail: ungar@cis.upenn.edu). This research was supported by a research contract to the University
of Pennsylvania and the University of California from the Intelligence
Advanced Research Projects Activity (IARPA) via the Department of
Interior National Business Center contract number D11PC20061. The
U.S. Government is authorized to reproduce and distribute reprints for
Government purposes notwithstanding any copyright annotation
thereon. Disclaimer: The views and conclusions expressed herein are
those of the authors and should not be interpreted as necessarily
representing the official policies or endorsements, either expressed
or implied, of IARPA, DoI/NBC, or the U.S. Government. The authors would also like to thank Don Moore for providing us with the weight dataset.}} 
\date{\vspace{-8.5ex}}

\begin{document}
\maketitle


\date{\vspace{-8.5ex}}

\begin{abstract}
\singlespace
Prediction polling is an increasingly popular form of crowdsourcing in
which multiple participants estimate the probability or magnitude of
some future event. These estimates are then aggregated into a single
forecast. Historically, randomness in scientific estimation has been
generally assumed to arise from unmeasured factors which are viewed as
measurement noise. However, when combining subjective estimates, heterogeneity
stemming from differences in the participants' information is often
more important than measurement noise. This paper formalizes
information diversity as an alternative source of such heterogeneity
and introduces a novel modeling framework that is particularly
well-suited for prediction polls. A practical specification of this
framework is proposed and applied to the task of aggregating
probability and point estimates from two real-world prediction
polls. In both cases our model outperforms standard
measurement-error-based aggregators, hence providing evidence in favor
of information diversity being the more important source of
heterogeneity.
\\
\\
\textit{Keywords:} Expert belief; Forecast heterogeneity; Judgmental forecasting; Model
averaging; Noise reduction
\end{abstract}


\section{INTRODUCTION}
Past literature has distinguished two types of polling: prediction and opinion polling. In broad terms, an opinion poll is a survey of public opinion, whereas a prediction poll involves multiple agents collectively predicting the value of some quantity of interest \citep{goel2010prediction, mellers2014psychological}. For instance, consider a presidential election poll. An opinion poll typically asks the voters who they will vote for. A prediction poll, on the other hand, could ask which candidate they think will win in their state. A liberal voter in a dominantly conservative state is likely to answer differently to these two questions. 
Even though opinion polls have been the dominant focus historically, prediction polls have become increasingly popular in the recent years, due to modern social and computer networks  that permit the collection of a large number of responses both from human and machine agents. This has given rise to crowdsourcing platforms, such as MTurk and Witkey, and many companies, such as Myriada, Lumenogic, and Inkling, that have managed to successfully  capitalize on the benefits of collective wisdom.



%
%
%

\begin{figure}[t!]
\captionsetup{width=0.48\textwidth}
\begin{minipage}[t]{.5\textwidth}
                \includegraphics[width=\textwidth]{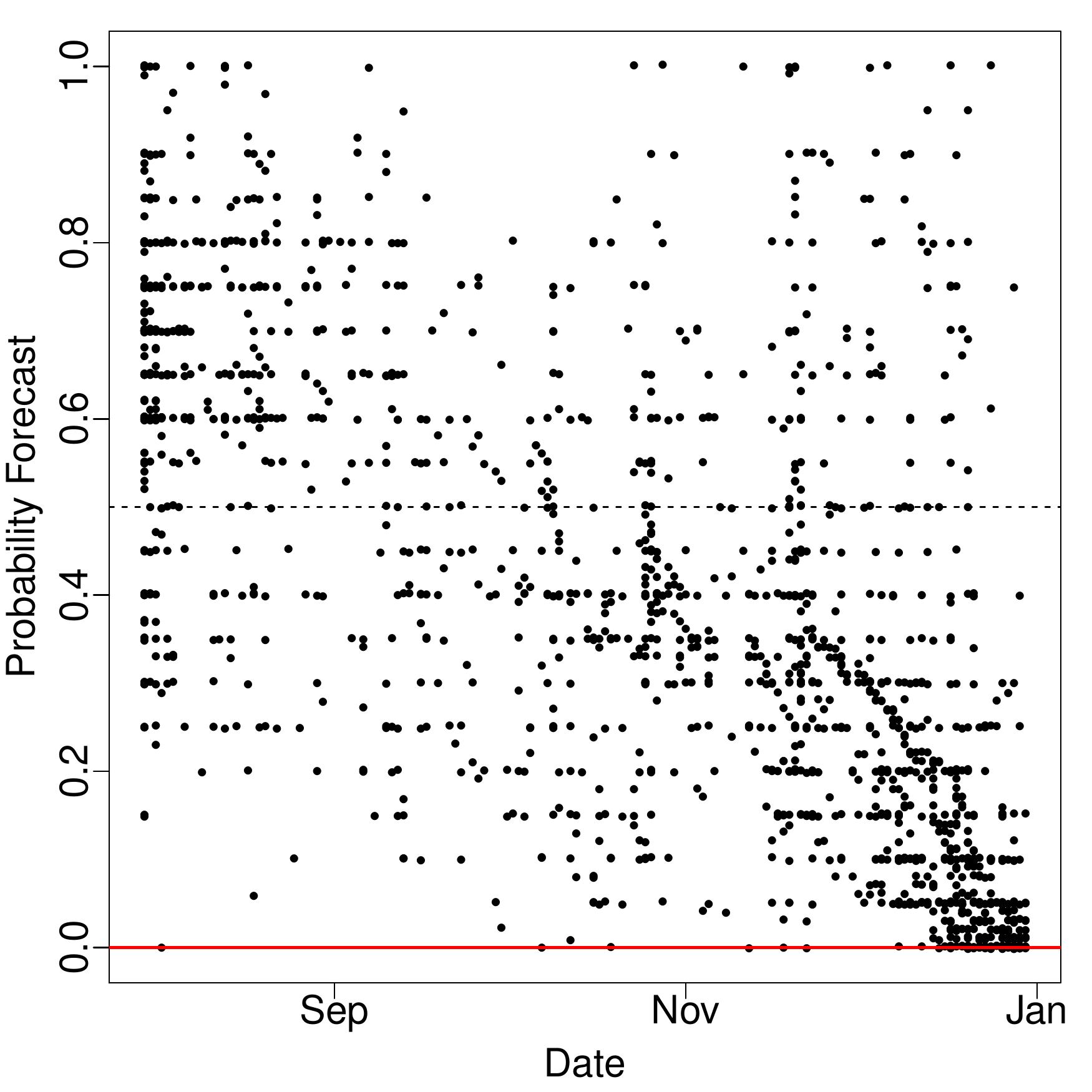}
                \caption{Probability forecasts of the event ``Will Moody's issue a new downgrade on the long-term ratings for any of the eight major French banks between 30 July 2012 and 31 December 2012?'' The points have been jittered slightly to make overlaps visible.}
                                \label{Example_prob}
\end{minipage}%
\begin{minipage}[t]{.5\textwidth}
                \includegraphics[width=\textwidth]{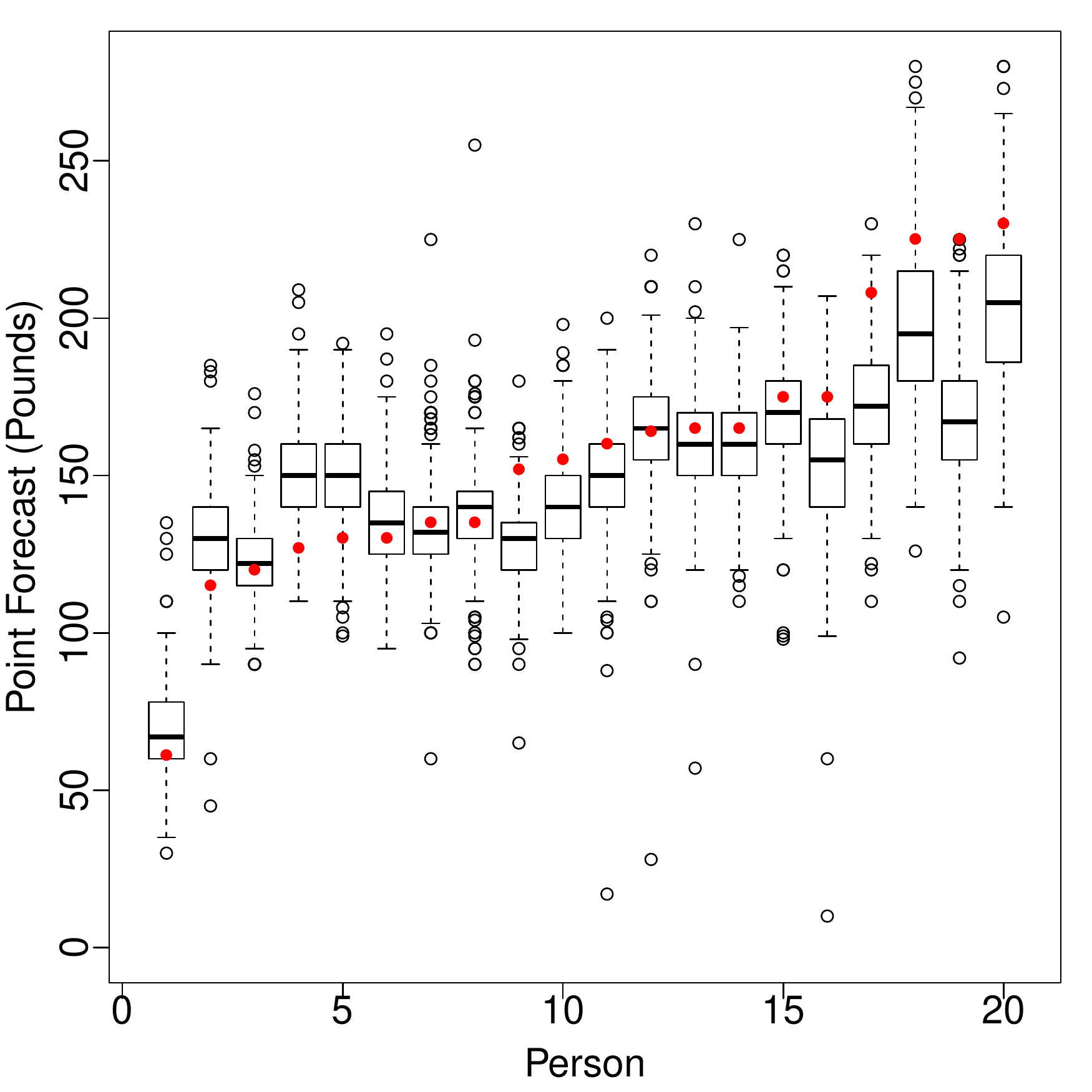}
                \caption{Point forecasts of the weights of $20$ different people. The boxplots have been sorted to increase in the true weights (red dots). Some extreme values were omitted for the sake of clarity.}
                                \label{Example_point}
        
\end{minipage}%
\end{figure}

This paper introduces statistical methodology designed specifically for the rapidly growing practice of prediction polling. The methods are illustrated on real-world data involving two common types of responses, namely probability and point forecasts. The probability forecasts were collected 
by the Good Judgment Project (GJP) (\citealt{ungar2012good, mellers2014psychological}) as a means to estimate the likelihoods of international political future events deemed important by the Intelligence Advanced Research Projects Activity (IARPA). 
Since its initiation in 2011, the project has recruited thousands of forecasters to make probability estimates and update them whenever they felt the likelihoods had changed. To illustrate, Figure \ref{Example_prob} shows the forecasts for one of these events. This example involves $522$ forecasters making a total of $1,669$ predictions between 30 July 2012 and 30 December 2012 when the event finally resolved as ``No'' (represented by the red line at $0.0$). 
In general, the forecasters reported updates very infrequently. Furthermore, not all forecasters made probability estimates for all the events, making the dataset very sparse. The point forecasts for our second application were collected by \cite{moore2008use} who recruited $416$ undergraduates from Carnegie Mellon University to guess the weights of $20$ people based on a series of pictures. This is an experimental setup where each participant was required to respond to all the questions, leading to a fully completed dataset. The responses are illustrated in Figure \ref{Example_point} that shows the boxplots of the forecasters' guesses for each of the $20$ people. The red dots represent the corresponding true weights.

Once the predictions have been collected, they are typically combined into a single consensus forecast for the sake of decision-making and improved accuracy. Unfortunately, this can be done in many different ways, and the final combination rule can largely determine the out-of-sample performance. The past literature distinguishes two broad approaches to forecast aggregation: empirical aggregation and model-based aggregation. Empirical aggregation is by far the more widely studied approach; see, e.g., stacking \citep{breiman1996stacked}, Bayes model averaging \citep{raftery1997bayesian}, linear opinion pools \citep{degroot1991optimal}, and extremizing aggregators \citep{Ranjan08, satopaa,satopaa2014probability}. All these methods are akin to machine learning in a sense that they first learn the aggregator based on a training set of past forecasts of known outcomes and then use that aggregator to combine future forecasts of unknown outcomes. 
 Unfortunately, in a prediction polling setup, constructing such a training set requires a lot of effort and time on behalf of the forecasters and the polling agent. Therefore a training set is often not available. Instead, the participants are typically handed a single questionnaire that simultaneously inquires about their predictions of one or more unknown outcomes. This leads to a dataset consisting only of forecasts, which means that empirical aggregation cannot be applied. 

Fortunately, model-based aggregation can be performed even when prior knowledge of outcomes is not available. This approach begins by proposing a plausible probability model for the source of heterogeneity among the forecasts, that is, for how and why the forecasts differ from the target outcome. Under this assumed forecast-outcome link, it is then possible to construct an optimal aggregator that can be applied directly to the forecasts without learning the aggregator first from a separate training set. Given this broad applicability, the current paper focuses only on the model-based approach. In particular, outcomes are not assumed available for aggregation at any point in the paper. Instead, aggregation is performed solely based on forecasts, leaving all empirical techniques well outside the scope of the paper.
%

Historically, potentially due to early forms of data collection, model-based aggregation has considered measurement error as the main source of forecast heterogeneity. This choice motivates aggregators with central tendency such as the (weighted) average, median, and so on. 
Intuitively, measurement error may be reasonable in modeling repeated estimates from a single instrument. However, it is unlikely to hold in prediction polling, where the estimates arise from multiple, often widely different sources. It is also known that a non-trivial weighted average is not the optimal aggregator (in terms of the expected quadratic and many other loss functions) under any joint distribution of the outcome and its (conditionally unbiased) forecasts \citep{dawid1995coherent,Ranjan08, satopaa2015combining}. This questions the role of measurement error in model-based aggregation and highlights the need for a different source of forecast heterogeneity.
 
 The main contribution of this paper is a new source of forecast heterogeneity, called \textit{information diversity}, that explains variation by differences in the information available to the forecasters and how they decide to use it. For instance, forecasters studying the same (or different) articles about a company may use separate parts of the information and hence report differing predictions on the company's future revenue. Such diversity forms the basis of a novel modeling framework known as the \textit{partial information framework}. Theory behind this framework was originally introduced for probability forecasts by \cite{satopaamodeling}; though their specification is somewhat restrictive for empirical applications. The current paper generalizes the framework beyond probability forecast and removes all unnecessary assumptions, leading to a new specification that is more appropriate for practical applications. 
This specification allows the decision-maker to build models for different types of forecast-outcome pairs, such as probability forecasts of binary events or point forecasts of real-valued outcomes. Each such model motivates and describes an explicit joint distribution for the target outcome and its forecasts. 
The optimal aggregator under this joint distribution is available and  
serves as a more principled model-based alternative to the usual (weighted) average or median.
The paper is structured as follows. Section \ref{PIF} first describes the partial information framework at its most general level and then introduces a practical specification of the framework. 
The section ends with a brief review of previous work on model-based aggregation. Section \ref{estimator} derives a general procedure that guides efficient estimation of the information structure among the forecasters. Section \ref{applications} illustrates on real-world data how specific models within the framework can be constructed and applied. In particular, the models are derived and evaluated on probability and point forecasts from the two prediction polls discussed above. Overall, the resulting partial information aggregators achieve a noticeable performance improvement over the common measurement-error-based aggregators, suggesting that  information diversity is the more appropriate model of forecast heterogeneity. 
 Finally, Section \ref{discussion} concludes with a summary and discussion of
future research.

\section{MODEL-BASED AGGREGATION}
\label{PIF}
\subsection{Bias and Noise}
Consider $N$ forecasters and suppose forecaster $j$ predicts $X_j$ for some quantity of interest $Y$. For instance, in our weight estimation example $Y$ is the true weight of a person and $X_j$ is the guess given by the $j$th undergraduate. In our probability forecasting application, on the other hand, $Y$ is binary, reflecting whether the event happens or not, and $X_j \in [0,1]$ is a probability forecast for its occurrence. This section, however, avoids such application specific choices and treats $Y$ and $X_j$ as generic random variables. In general, prediction $X_j$ is nothing but an estimator of $Y$. Therefore, as is the case with all estimators, 
its deviation from the truth can be broken down into two components: bias and noise. On the
theoretical level, these two components can be separated and hence are often addressed by different mechanisms. This suggests 
 a two-step approach to forecast aggregation: 
i) eliminate any bias in the forecasts, and ii) combine the unbiased forecasts.

Historically, bias in human judgment has been extensively studied in the psychology literature (for reviews, see \citealt{lichtenstein1977calibration, yates1990judgment, keren1991calibration}). This bias often exhibits well-known patterns (see, e.g., the easy-hard effect in \citealt{lichtenstein1977those, juslin1993explanation}), and many authors have proposed both cognitive and motivational models to explain it \citep{koriat1980reasons, kruglanski1990motivations, soll1996determinants, moore2008trouble}. These models and other results in this popular area of research suggest ways for ex-ante bias reduction. 
%
Such techniques, however, are not in the scope of this paper. Instead, the focus here is on noise reduction and hence specifically on developing methodology for the second step in the overall process of forecast aggregation. 
In particular, Section \ref{generalPIF} describes our new framework for modeling the noise component. This is then compared in Section \ref{otherModels} to previous noise models.
 These models make different assumptions about the way the unbiased forecasts relate to the target outcome and hence motivate very different classes of model-based aggregators. 
%


\subsection{Partial Information Framework}
\label{generalPIF}
\subsubsection{General Framework}
\label{context}

\noindent
The partial information framework assumes that $Y$ and $X_j$ are measurable under some common probability space $(\Omega, \F , \P)$.  The probability measure $\P$ provides a non-informative yet proper  prior on $Y$ and  reflects the \textit{basic information} known to all forecasters. 
Such a prior has been discussed extensively in the economics and game theory literature where it is usually known as the \textit{common prior}. 
Even though this is a substantive assumption in the framework, specifying a prior distribution cannot be avoided as
long as the model depends on a probability space. This includes essentially any probability
model for forecast aggregation. How the prior is incorporated depends on the problem context: it can be chosen explicitly by the decision-maker, computed based on past observations of $Y$, or estimated directly from the forecasts. 


The principal $\sigma$-field $\F$ can be interpreted as all the possible information that can be known about $Y$. 
On top of the basic information reflected in the prior, the $j$th forecaster uses some personal partial information set $\F_j \subseteq \F$ and 
predicts $X_j = \E(Y \| \F_j)$.
Therefore $\F_i \neq \F_j$ if $X_i \neq X_j$, and forecast heterogeneity stems purely from \textit{information diversity}.  Note, however, that if forecaster $j$ uses a simple rule, $\F_j$ may not be the full $\sigma$-field of information available to the forecaster but rather a smaller $\sigma$-field corresponding to the information used by the rule. Furthermore, if two forecasters have access to the same $\sigma$-field, they may decide to use different sub-$\sigma$-fields, leading to different predictions. 
This is particularly salient in our weight estimation example where each forecaster has access to the exact same information, namely the picture of the person, but can choose to use different subsets of this information. 
Therefore,
information diversity does not only arise from differences in the available information, but also from how the forecasters decide to use it. This general point of view	 was motivated in \cite{satopaamodeling} with simple examples that illustrate how the optimal aggregate is not well-defined without assumptions on the information structure among the forecasters.

 \cite{satopaamodeling}  also show that $X_j = \E(Y \| \F_j)$ is precisely the same as having a calibrated (sometimes also known as reliable) forecast, that is, $X_j = \E(Y | X_j)$. 
Therefore the form $X_j = \E(Y | \F_j)$  arises directly from the existence of an underlying probability model and calibration. 
Overall, calibration $X_j = \E(Y | X_j)$ has been widely discussed in the  statistical and meteorological forecasting literature (see, e.g., \citealt{dawid1995coherent, Ranjan08, jolliffe2012forecast}), with traces at least as far back as \citet{murphy1987general}. Given that the condition $X_j = \E(Y | X_j)$  depends on the probability measure $\P$, it should be referred to as $\P$-calibration when the choice of the probability measure needs to be emphasized. This dependency shows the main conceptual difference between $\P$-calibration and the notion of empirical calibration (\citealt{dawid1982well, foster1998asymptotic}; and many others). However, as was pointed out by \citet{dawid1995coherent}, these two notions can be expressed in formally identical terms by letting $\P$ represent the limiting joint distribution of the forecast-outcome pairs.

In practice researchers have discovered many calibrated subpopulations of experts, such as meteorologists \citep{murphy1977can, murphy1977reliability}, experienced tournament bridge players \citep{keren1987facing}, and bookmakers \citep{dowie1976efficiency}. Generally, calibration can be improved through team collaboration, training, tracking \citep{mellers2014psychological}, performance feedback \citep{murphy1984impacts}, representative sampling of target events \citep{gigerenzer1991probabilistic, juslin1993explanation}, or by evaluating the forecasters' performance under a 
loss function that is minimized by the conditional expectation of $Y$, given the forecaster's information
%
 \citep{banerjee2005optimality}.
If one is nonetheless left with uncalibrated forecasts, they can be calibrated ex-ante as follows. First, consider some (possibly uncalibrated) forecasts $\tilde{\X} = (\tilde{X}_1, \dots, \tilde{X}_N)'$ defined on $(\Omega, \F)$. Choose some distribution $\Qp$ for $(Y, \tilde{\X})$. For instance, \citealt{dawid1995coherent}  suggest first choosing a distribution $\Qp$ for $\tilde{\X}$ and then setting  $\Qp(Y, \tilde{\X}) = \Psi(\tilde{\X}) \Qp( \tilde{\X})$, where
 $\Psi$ is an arbitrary aggregator (such as the average of probability forecasts of a binary event) acting as $ \Qp(Y| \tilde{\X})$.
%
Alternatively, one may search for an appropriate $\Qp$ in the large literature of quantitative psychology.  Regardless how $\Qp$ is constructed, however, the calibrated version of $\tilde{X}_j$ is  
 $\E_\Qp(Y|\tilde{X}_j)$.  This forecast is $\Qp$-calibrated and can be written as $\E_\Qp(Y | \F_j)$, where $\F_j = \sigma(\E_\Qp(Y|\tilde{X}_j))$ is the $\sigma$-field generated by $\E_\Qp(Y|\tilde{X}_j)$. Intuitively, calibrating is equivalent to replacing forecast $x$ by $\E_\Qp(Y | \tilde{X}_j = x)$ for all possible values $x \in \supp(\tilde{X}_j)$. Perhaps, however, one does not want to work under this particular model. To accommodate alternative models (such as the Gaussian model described in Section \ref{gaussian}), the next proposition shows how $\Qp$-calibrated forecasts can be transformed into forecasts that are calibrated under some other probability measure $\P$. All the proofs are deferred to Appendix A.
 \begin{proposition}
\label{modelTransformation}
Consider a probability measure $\P$ such that $\P \ll \Qp$. Let $\frac{d\P}{d\Qp}$ denote the Radon-Nikodym derivative of  $\P$ with respect to $\Qp$. The forecasts under the new model $\P$ are then given by the transformation $\E_\P(Y|\F_j) = \E_\Qp \left( \frac{d\P}{d\Qp} Y \big| \F_j \right) \big/ \E_\Qp\left(\frac{d\P}{d\Qp}  \big| \F_j \right)$, where $\F_j = \sigma(\E_\Qp(Y|\tilde{X}_j))$.
\end{proposition}
%
%
This shows that uncalibrated forecasts from ``non-experts'' can be calibrated as long as one agrees on some joint distribution for the target outcome and its forecasts. While such constructs certainly deserve further analysis, they are not in the scope of this paper and hence are left for future work. Therefore, from now on, the forecasts are assumed to be calibrated. Note, however, that in general the forecasts should satisfy some minimal performance criterion; simply aggregating entirely arbitrary forecasts is hardly going to lead to improved forecasting accuracy. To this end, \citet{foster1998asymptotic} analyze probability forecasts and state that ``calibration does seem to be an appealing minimal property that any probability forecast should satisfy.'' They show that one needs to know almost nothing about the  outcomes in order to be calibrated. Thus, in theory, calibration can be achieved very easily and overall seems like an appropriate base assumption for developing a general theory of forecast aggregation.

Given that the partial information framework generates all forecast variation from information diversity, it is important to understand the extent to which the forecasters' partial information sets can be measured in practice. First, note that, for the purposes of aggregation, any available information discarded by a forecaster may as well not exist because information comes to the aggregator only through the forecasts. Therefore it is not in any way restrictive to assume that $\F_j = \sigma(X_j)$. Second,  the following proposition describes observable measures for the amount of information in each forecast and for the amount of information overlap between any two forecasts. 

 \begin{proposition}
\label{covstr}
If $\F_j = \sigma(X_j)$ such that $\E(Y|\F_j) = \E(Y|X_j)  = X_j$ for all $j =1, \dots, N$, then the following holds.
\begin{enumerate}[i)] 
\item Forecasts are marginally consistent: $\E(Y) = \E(X_j)$. \label{first}
\item Variance increases in information: $\Var(X_i) \leq \Var(X_j)$ if $\F_i \subseteq \F_j$. Given that $Y = \E(Y | \F)$, the variances of the forecasts are upper bounded as $\Var(X_j) \leq \Var(Y)$ for all $j = 1, \dots, N$. \label{third}
\item $\Cov(X_j, X_i) = \Var(X_i)$ if $\F_i \subseteq \F_j$. Again, expressing $Y = \E(Y | \F)$ implies that $\Cov(X_j, Y) = \Var(X_j)$ for all $j = 1, \dots, N$.
   \label{second}
\end{enumerate}
\end{proposition}

This proposition is important for multiple reasons. First, item \ref{first}) provides guidance in estimating the prior mean of $Y$ from the observed forecasts. Second, item \ref{third}) shows that $\Var(X_j)$ quantifies the amount of information used by forecaster $j$. In particular, $\Var(X_j)$ increases to $\Var(Y)$ as forecaster $j$  learns and becomes more informed.  Therefore increased variance reflects more information and is deemed helpful. This is a clear contrast to the standard statistical models that often regard higher variance as increased noise and hence harmful. 
The covariance $\Cov (X_{i}, X_{j})$, on the other hand, can be interpreted as the amount of information overlap between forecasters $i$ and $j$. Given that being non-negatively correlated is not generally transitive \citep{langford2001property}, these covariances are not necessarily non-negative even though all forecasts are non-negatively correlated with the outcome. Such negatively correlated forecasts can arise in a real-world setting. For instance, consider two forecasters who see voting preferences of two different
sub-populations that are politically opposed to each other.  Each
individually is a weak predictor of the total vote on any given issue,
but they are negatively correlated because of the likelihood that
these two blocks will largely oppose each other.
%

Third and finally, item \ref{second})  shows that the covariance matrix $\bSigma_X$ of the $X_j$s extends to the unknown $Y$ as follows:
\begin{align}
\Cov\left((Y, X_1, \dots, X_N)'\right) &=  \left( \begin{matrix} 
 \Var(Y)  & \diag(\bSigma_X)'  \\
\diag(\bSigma_X) & \bSigma_X \\
\end{matrix} \right), \label{cov_str}
\end{align}
where $\diag(\bSigma_X)$ denotes the diagonal of $\bSigma_X$.
%
This is the key to regressing $Y$ on the $X_j$s without a separate training set of past forecasts of known outcomes. The resulting estimator, called the \textit{revealed aggregator}, is $$X'' := \E(Y | X_1, \dots, X_N) = \E \left(Y \|
\F''\right),$$ where $\F'' := \sigma(X_1, \dots, X_N)$ is the $\sigma$-field generated (or information revealed) by the
$X_j$s. The revealed aggregator uses all the information that is available in the forecasts and hence is the optimal aggregator under the distribution of $(Y, X_1, \dots, X_N)$.
 To make this precise, 
consider a scoring rule $S(x,y)$ that represents the loss of predicting $x$ when the outcome is $y$. A scoring rule is said to be consistent for the mean of $Y$ if $\E_{Y}[S(\E_{Y}(Y), Y)] \leq \E_{Y}[S(x, Y)]$ for all $x \in \R$. \citet{savage1971elicitation} showed, subject to weak regularity conditions, that all such scoring rules can be written in the form
\begin{align}
S(x,y) &= \phi(y) -\phi(x) - \phi'(x)(y-x), \label{lossForm}
\end{align}
where $\phi$ is a convex function with subgradient $\phi'$. An important special case is the quadratic loss $S(x,y) = (x-y)^2$ that arises when $\phi(x) = x^2$. Now, if an aggregator is defined as any random variable $X \in \sigma(X_1, \dots, X_N)$, then $X''$ is an aggregator that minimizes expectation of any scoring rule $S$ of the form (\ref{lossForm}):
\begin{align*}
\E[S(X,Y)] &= \E_{X_1, \dots, X_N}\{ \E_{Y | X_1, \dots, X_N}[S(X,Y) ]\}\\
&\geq \E_{X_1, \dots, X_N}\{ \E_{Y | X_1, \dots, X_N}[S(X'',Y) ]\}\\
&= \E[S(X'',Y)].
\end{align*}
\citet{Ranjan08} showed a similar results for probability forecasts. For these reasons, $X''$ is considered the relevant aggregator under each specific instance of the framework. The next section shows how this aggregator can be captured in practice.

\subsubsection{Gaussian Partial Information Model}
\label{gaussian}
Even though the general framework is convenient for theoretical analysis, it is clearly too abstract for practical applications. Fortunately, applying the framework in practice only requires one extra assumption, namely the choice of a parametric family for the distribution of 
$(Y, X_1, \dots, X_N)$. 
One approach is to refer to Proposition \ref{covstr} and choose a family that is parametrized in terms of the first two joint moments. This points at the multivariate Gaussian distribution that is a typical starting point in developing statistical methodology and often provides the cleanest entry into the issues at hand. 
The Gaussian distribution is also the most common choice for modeling measurement error. This is typically motivated by assuming the terms to represent sums of a large number of independent sources of error. The central limit theorem then gives a natural motivation for the Gaussian distribution. A similar argument can be made under the partial information framework.
First, consider some pieces of information. Each piece either has a positive or negative impact and hence respectively either increases or decreases $Y$. The total sum (integral) of these pieces determines the value of $Y$. Each forecaster, however, only observes the sum of some subset of them. Based on this sum, the forecaster makes an estimate of $Y$. 
If
the pieces are independent
and have small tails, then the joint distribution
of the forecasters' observations will be asymptotically
Gaussian.  Given that the number of information pieces in a real-world setup is likely to be large, it makes sense
to model the forecasters' observations as jointly Gaussian. Of course, other distributions, such as the multivariate $t$-distribution, are possible. At this point, however, such alternative specifications are best left for future work.
%
%

The model variables $(Y, X_1, \dots, X_N)$ can be modeled directly with a Gaussian distribution as long as they are all real-valued. In many applications, however, $Y$ and $X_j$ may not be supported on the whole real line. For instance, the aforementioned Good Judgment Project  collected probability forecasts of binary events. In this case,  $X_j \in [0,1]$ and $Y \in \{0,1\}$. Fortunately, different types of outcome-forecast pairs can be easily addressed by borrowing from the theory of generalized linear models \citep{mccullagh1989generalized} and utilizing a \textit{link function}. The result is a close yet widely applicable specification called the \textit{Gaussian partial information model}. This model begins by introducing $N+1$ \textit{information variables} that follow a multivariate Gaussian distribution with the covariance pattern (\ref{cov_str}):
\begin{align}
\left(\begin{matrix} Z_0 \\ Z_{1}\\ \vdots \\ Z_{N} \end{matrix}\right) &\sim \mathcal{N}_{N+1}\left( 
 \boldsymbol{0}, \left(\begin{matrix} 
1 & \diag(\bSigma)'\\
\diag(\bSigma) &\bSigma\\
 \end{matrix}\right) 
 :=
 \left(\begin{array}{c | c c cc }
1 & \delta_1 & \delta_2 & \dots & \delta_N  \\ \hline
\delta_1 & \delta_1 &\rho_{1,2} & \dots & \rho_{1,N}   \\ 
\delta_2 & \rho_{2,1} & \delta_2 & \dots & \rho_{2,N}  \\ 
\vdots & \vdots & \vdots & \ddots & \vdots  \\ 
\delta_N & \rho_{N,1} & \rho_{N,2} & \dots & \delta_N\\ 
 \end{array}\right)\right).  \label{NExperts}
\end{align}
This distribution supports the Gaussian model similarly to the way the ordinary linear regression supports  the class of generalized linear models. 
In particular, the information variables transform into the outcome and forecasts via an application-specific link function $g(\cdot)$; that is, $Y = g(Z_0)$ and $X_j = \E(Y | Z_j) = \E(g(Z_0) | Z_j)$. Given that $Z_0$ fully determines $Y$, it is sufficient for all information that can be known about $Y$. The remaining variables $Z_1, \dots, Z_N$, on the other hand, summarize the forecasters' partial information. To make this more concrete, consider our two real-world applications. 
For probability forecasts of a binary event a reasonable link function $g(\cdot)$ is the indicator function $\one_{A}$, where $A = \{Z_0 > t\}$ for some threshold value $t \in \R$. For real-valued $X_j$ and $Y$, on the other hand, a reasonable choice is the reverse standardizing function $g(Z_0) = \sigma_0 Z_0 + \mu_0$, where $\mu_0$ and $\sigma_0$ are the prior mean and standard deviation of $Y$, respectively. In general, it makes sense to have $g(\cdot)$ map from the real-numbers to the support of $Y$ such that $Y$ has the correct prior $\P(Y)$. 


%

%
Overall, this model can be considered as a close yet practical specification of the general framework. After all, it only adds on the assumption of Gaussianity. This extra assumption, however, is enough to allow the construction of the revealed aggregator $X'' = \E(Y | Z_1, \dots, Z_N)$. For $X''$ and also $X_j$ the conditional expectations can be often computed via the following conditional distributions:
\begin{align*}
 Z_0 | Z_j &\sim \mathcal{N}\left(Z_j, 1-\delta_j\right) \text{ and }\\
Z_0 | \Z &\sim \mathcal{N}\left(\diag(\bSigma)' \bSigma^{-1} \Z, 1-\diag(\bSigma)' \bSigma^{-1}\diag(\bSigma)\right),
\end{align*}
where $\Z = (Z_1, \dots, Z_N)'$. For instance, if both $X_j$ and $Y$ are real-valued, then $X_j = \sigma_0 Z_j + \mu_0$ and $X'' = \diag(\bSigma)'\bSigma^{-1}(\X-\mu_0\one_N) + \mu_0$, where $\X = (X_1, \dots, X_N)'$. These conditional distributions arise directly from the well-known conditional distributions of the multivariate Gaussian distribution (see, e.g., \citealt{ravishanker2001first}). 

\subsection{Previous Work on Model-Based Aggregation}
\label{otherModels}

\subsubsection{Interpreted Signal Framework}

The {\em interpreted signal
framework} is a behavioral model that assumes different 
predictions to arise from differing interpretation procedures \citep{hong2009interpreted}.
For
example, consider two forecasters who visit a company and predict its future revenue. One forecaster may carefully examine the company's technological status while the other pays closer attention to what the managers say.  Even though the
forecasters receive and possibly even use the exact same information, they may interpret it
differently and hence end up reporting different forecasts. Therefore forecast heterogeneity is assumed to stem from ``cognitive
diversity''.

This is a very reasonable model and hence has been used in various forms to simulate and illustrate theory about expert behavior (see, e.g., \citealt{broomell2009experts, parunak2013characterizing}).  
%
Consequently, previous authors have constructed many highly specialized toy models of interpreted forecasts. For instance,
\cite{dawid1995coherent} construct simple models of two forecasts to support their discussion on coherent forecast aggregation; \cite{Ranjan08} use one of these models to simulate calibrated forecasts; and  \cite{Bacco} introduce a model for two forecasters whose (interpreted) log-odds predictions follow a joint Gaussian distribution. Unfortunately, their model is very narrow due to its detailed
assumptions and extensive computations. 
Furthermore, it is not clear how the model can be used in practice or extended to $N$ forecasters. 
All in all, it seems that successful previous applications of the interpreted signal framework have used it as a basis for illustrating theory instead of actually aiming to model real-world forecasts. 
In this respect, the framework has remained relatively abstract. 

Our partial information framework, however, formalizes the intuition behind it, allows quantitative predictions, and provides a flexible construction for modeling many different forecasting setups. Overall, the framework is very general and, in fact, encompasses all the other authors' models mentioned above as different sub-cases. Unlike the Gaussian model, however, these models make many restrictive assumptions in addition to just choosing a parametric family. Even though the general partial information framework, as described in Section \ref{generalPIF}, does not allow the forecasters to interpret information differently and hence does not capture all aspects of the interpreted signal framework, personal interpretations can be easily introduced by associating forecaster $j$ with a probability measure $\P_j$  that describes that forecaster's interpretation of information. If $\E_j$ denotes the expectation under $\P_j$, then it is possible that $X_i = \E_i(Y | \F_i) \neq X_j = \E_j(Y | \F_j)$ even if $\F_i = \F_j$. In practice, however, eliciting the details of each $\P_j$ is hardly possible. Therefore, to keep the model tractable, it is convenient to assume a common interpretation $\P_j = \P$ for all $j = 1, \dots, N$.

\subsubsection{Measurement Error Framework}
In the absence of a quantitative
interpreted signal model, prior applications have typically explained forecast heterogeneity  with standard statistical models. These models are different formalizations of  the \textit{measurement error framework} that generates forecast heterogeneity purely from a probability distribution. More specifically, this framework assumes a ``true'' (possibly transformed) forecast
$\theta$, which can be interpreted as the prediction made by an ideal forecaster. The forecasters then somehow measure $\theta$ with mean-zero idiosyncratic error. For instance, in our probability forecasting application one possible measurement error model is 
\begin{align}
Y &\sim \text{Bernoulli}(\theta), \nonumber\\
\logit (X_j) &= \logit(\theta) + e_j, \text{ and} \label{MsrModel}\\
e_j &\stackrel{i.i.d.}{\sim} \mathcal{N}(0,\sigma^2) \text{ for all }j = 1, \dots, N, \nonumber
\end{align}
where $\logit(x) = \log(x/(1-x))$ is the log-odds operator. Given that the errors are generally assumed to  have mean zero, measurement error forecasts are unbiased estimates of $\theta$, that is, $\E(X_j | \theta) = \theta$. Observe that this is not the same as assuming calibration $\E(Y|X_j) = X_j$. Therefore an unbiased
estimation model is very different from a calibrated model.
This distinction is further emphasized by the fact that $X''$ never reduces to a (non-trivial) weighted average of the forecasts \citep{satopaa2015combining}. Given that the measurement-error aggregators are often different types of weighted averages, measurement error and information diversity are not only philosophically different but they also require very different aggregators. 



%


Example (\ref{MsrModel}) illustrates the main advantages of the measurement error framework: simplicity and familiarity. 
Unfortunately, there are a number of disadvantages. First, measurement-error aggregators estimate $\theta$ instead of the realized value of the random variable $Y$. 
For this reason, these aggregators often do not satisfy even the minimal performance requirements. For instance, a non-trivial weighted average of calibrated forecasts is necessarily both uncalibrated and under-confident \citep{Ranjan08, satopaa2015combining}. 
Second, the standard assumption of conditional independence of the observations forces a specific and highly
unrealistic structure on interpreted forecasts \citep{hong2009interpreted}. Measurement-error aggregators also cannot leave
the convex hull of the individual forecasts, which further contradicts the
interpreted signal framework \citep{parunak2013characterizing} and can be easily seen to result in poor empirical performance on many datasets.
 Third,  the
underlying model is rather implausible. Relying on a true forecast $\theta$ invites philosophical debate, and even if one assumes the existence of such a value, it is difficult to believe that the forecasters are actually seeing it with independent noise. Therefore, whereas the
interpreted signal framework proposes a plausible micro-level explanation, the
measurement error model does not; at best, it forces us to imagine a group of forecasters who apply the same
procedures to the same data but with numerous small mistakes.

\section{MODEL ESTIMATION}
\label{estimator}
This section describes methodology for estimating the \textit{information structure} $\bSigma$. Even though $\bSigma$ is mostly used for aggregation, it also describes the information among the forecasters (see end of Section \ref{context}) and hence should be of interest to decision analysts, psychologists, and the broader community studying  collective problem solving.
Unfortunately, estimating $\bSigma$ in full generality based on a single prediction per forecaster is difficult. Therefore, to facilitate model estimation, the forecasters are assumed to predict $K \geq 2$ related events. For instance, in our second application $416$ undergraduates guessed the weights of $20$ people. This yielded a $20 \times 416$ matrix that was then used to estimate $\bSigma$.



\subsection{General Estimation Problem}
\label{generalEstimation}

Denote the outcome of the $k$th event with $Y_k$ and the $j$th forecaster's prediction for this outcome with $X_{jk}$. For the sake of generality, this section does not assume any particular link function but instead operates directly with the corresponding information variables, denoted with $Z_{jk}$. In practice, the forecasts $X_{jk}$ can be often transformed into $Z_{jk}$ at least approximately. This is illustrated in Section \ref{applications}. Recall that aggregation cannot access to the outcomes $\{Y_1, \dots, Y_K\}$ or their corresponding information variables $\{Z_{01}, \dots, Z_{0K}\}$. Instead, $\bSigma$ is estimated  
only based on $\{\Z_1, \dots, \Z_K\}$, where the vector $\Z_k = (Z_{1k}, \dots, Z_{Nk})'$ collects the forecasters' information about the $k$th event. 
  
This estimation must respect the covariance pattern (\ref{NExperts}). More specifically, if $\mathcal{S}_{+}^N$ denotes the set of $N \times N$ symmetric positive semidefinite matrices  and 
\begin{align*}
h(\M)  := 
\left(\begin{matrix} 
1 & \diag(\M)' \\
\diag(\M) &\M\\
 \end{matrix}\right)
\end{align*}
for some symmetric matrix $\M$,
then the final estimate must satisfy the condition $h(\bSigma) \in \mathcal{S}_{+}^{N+1}$.
Intuitively, this is satisfied if there exists a random variable $Y$ for which the forecasts $X_j$ are jointly calibrated. In terms of information, this means that it is physically possible to allocate information about $Y$ among the $N$ forecasters in the manner described by $\bSigma$. Therefore the condition is named \textit{information coherence}. 

Unfortunately, simply finding an accurate estimate of $\bSigma$ does not guarantee precise aggregation. To see this, recall from Section \ref{gaussian} that $\E(Z_{0k} | \Z_k)  = \diag(\bSigma)' \bSigma^{-1} \Z_k$. 
This term is generally found in the revealed aggregator and hence deserves careful treatment. 
 Re-express the term as $\v' \Z_k$, where $\v$ is the solution to $\diag(\bSigma) = \bSigma \v$. The rate at which the solution changes with respect to a change in $\diag(\bSigma)$ depends on the condition number $\cond(\bSigma) := \lambda_{max}(\bSigma) / \lambda_{min}(\bSigma)$, i.e., the ratio between the maximum and minimum eigenvalues of $\bSigma$. If the condition number is very large, a small error in $\diag(\bSigma)$ can cause a large error in $\v$. If the condition number is small, $\bSigma$ is called \textit{well-conditioned} and error in $\v$ will not be much larger than the error in $\diag(\bSigma)$. Thus, to prevent estimation error from being amplified during aggregation, the estimation procedure should require $\cond(\bSigma) \leq \kappa$ for a given threshold $\kappa \geq 1$.

 
%



This all gives the following general estimation problem:
\begin{align}
\label{orig_problem}
\begin{split}
\text{minimize } &  f_0 \left(\bSigma, \{\Z_1, \dots, \Z_k\}\right) \\
\text{subject to } & h(\bSigma) \in \mathcal{S}_{+}^{N+1}, \text{ and}\\
& \cond(\bSigma) \leq \kappa,
\end{split}
\end{align}
where $f_0$ is some objective function. The feasible region defined by the two constraints is convex. Therefore,
if $f_0$ is convex in $\bSigma$, expression (\ref{orig_problem}) is a convex optimization problem. 
Typically the global optimum to such a problem can be found very efficiently. Problem (\ref{orig_problem}), however, involves $\binom{N+1}{2}$ variables. Therefore it can be solved efficiently with standard optimization techniques, such as the interior point methods, as long as the number of variables is not too large, say, not more than 1,000. Unfortunately, this means that the procedure cannot be applied to prediction polls with more than about $N = 45$ forecasters. This is very limiting as many prediction polls involve hundreds of forecasters. For instance, our two real-world applications involve $100$ and $416$ forecasters. Fortunately,  by choosing the loss function carefully one can perform dimension reduction and estimate $\bSigma$ under a much larger $N$. This is illustrated in the following subsections.

\subsection{Maximum Likelihood Estimator}
\label{mle}

Under the Gaussian model the information structure $\bSigma$ is a parameter of an explicit likelihood. Therefore estimation naturally begins with the maximum likelihood approach (MLE).  Unfortunately, the Gaussian likelihood is not convex in $\bSigma$. Consequently, only a locally optimal solution is guaranteed with standard optimization techniques. 
Furthermore, it is not clear whether the dimension of this form can be reduced. \cite{won2006maximum} discuss the MLE under a condition number constraint. They are able to transform the original problem with $\binom{N+1}{2}$ variables to an equivalent problem with only $N$ variables, namely the eigenvalues of $\bSigma$. This transformation, however, requires an orthogonally invariant problem. Given that the constraint $h(\bSigma) \in \mathcal{S}_{+}^{N+1}$ is not orthogonally invariant, the same dimension-reduction technique cannot be applied. Instead, the MLE must be computed with the $\binom{N+1}{2}$ variables, making estimation slow for small $N$ and undoable even for moderately large $N$. For these reasons the MLE is not discussed further in this paper.

\subsection{Least Squares Estimator}
\label{lse}






Past literature has discussed many simple covariance estimators that can be applied efficiently to large amounts of data. Unfortunately, these estimators are not guaranteed to satisfy the conditions in (\ref{orig_problem}). This section introduces a correctional procedure that inputs any covariance estimator $\SS$ and modifies it minimally such that the end result satisfies the conditions in (\ref{orig_problem}). More specifically, $\SS$ is projected onto the feasible region.  
 This approach, sometimes known as the least squares approach (LSE), 
motivates a convex loss function that guarantees a globally optimal solution and facilitates dimension reduction. Most importantly, however, it provides a general tool for estimating $\bSigma$, regardless whether one is working with a Gaussian model or possibly some future non-Gaussian model. 

From the computational perspective, it is more convenient to project $h(\SS)$ instead of $\SS$.  Even though this could be done under many different norms, for the sake of simplicity, this paper only considers the squared Frobenius norm $||\M||_F^2 = \Tr(\M'\M)$, where  $\Tr(\cdot)$ is the trace operator.  The LSE is then given by $h^{-1}(\bOmega)$, i.e., $\bOmega$ without the first row and column, where $\bOmega$ is the solution to
\begin{align}
 \label{lse_problem}
 \begin{split}
\text{minimize } &  || \bOmega - h(\boldsymbol{S}) ||_F^2\\
\text{subject to } & \bOmega \in \mathcal{S}_{+}^{N+1},\\
& \cond(\bOmega) \leq \kappa, \text{ and }\\
& \Tr(\A_j \bOmega) = b_j, \hspace{2em} (j = 1, \dots, N+1).
\end{split}
\end{align}
Both $\A_j$ and $b_j$ are constants defined to maintain the covariance pattern (\ref{NExperts}). 
More specifically, if $\e_j$ denotes the $j$th standard basis vector of length $N+1$, then 
\begin{align*}
b_1 &= 1 \text{, } \A_1 = \e_1 \e_1' \text{, and}\\
b_j &= 0 \text{, } \A_j = \e_j\e_j' - 0.5(\e_1\e_j' + \e_j\e_1') \text{ for } j = 2, \dots, N+1.
\end{align*}
%
If $\bOmega$ satisfies the other two conditions, namely $\bOmega \in  \mathcal{S}_{+}^{N+1}$ and $ \cond(\bOmega) \leq \kappa$, then $\bSigma = h^{-1}(\bOmega)$ also satisfies them. This follows from the fact that $\bSigma$ is a principal sub-matrix of $\bOmega$. Therefore $\bOmega \in \mathcal{S}_{+}^{N+1}$ implies $\bSigma \in \mathcal{S}_{+}^{N}$. Furthermore, Cauchy's interlace theorem (see, e.g., \citealt{hwang2004cauchy}) states that $\lambda_{min}(\bOmega) \leq \lambda_{min}(\bSigma)$ and $\lambda_{max}(\bSigma) \leq \lambda_{max}(\bOmega)$ such that $\cond(\bSigma) \leq \cond(\bOmega) \leq \kappa$. Of course, requiring $\cond(\bOmega) \leq \kappa$ instead of $\cond(\bSigma) \leq \kappa$ shrinks the region of feasible $\bSigma$s. 
 At this point, however, the exact value of $\kappa$ is arbitrary and merely serves to control $\cond(\bSigma)$.  
 Section \ref{condSelection} introduces a procedure for choosing $\kappa$ from the data.
  Under such an adaptive procedure, problem (\ref{lse_problem}) can be considered equivalent to directly projecting $\SS$ onto the feasible region. 


The first step towards solving (\ref{lse_problem}) is to express the feasible region as an intersection of the following two sets: 
\begin{align*}
\mathcal{C}_{sd} &= \left\{\bOmega : \bOmega \in \mathcal{S}_{+}^{N+1}, \cond(\bOmega) \leq \kappa\right\}, \text{ and }\\
\mathcal{C}_{lin} &= \left\{\bOmega : \Tr(\A_j \bOmega) = b_j, j = 1, \dots, N+1\right\}.
\end{align*}
Given that both of these sets are convex, projecting onto their intersection can be computed with the Directional Alternating Projection Algorithm \citep{gubin1967method}. This method makes progress by repeatedly projecting onto the sets $\mathcal{C}_{sd}$ and $\mathcal{C}_{lin}$. Consequently, it is efficient only if projecting onto each of the individual sets is fast. Fortunately, as will be shown next, this turns out to be the case.

%
%




First, projecting an $(N+1) \times (N+1)$ symmetric matrix $\M = \{m_{ij}\}$  onto $\mathcal{C}_{lin}$ is a linear map. To make this more specific, let $\m = \vec(\M)$ be a column-wise vectorization of $\M$. 
If $\A$ is a matrix with the $j$th row equal to $\vec(\A_j)$, the  linear constraints in (\ref{lse_problem}) can be expressed as $\A\m =  \boldsymbol{e}_1$. Then, the projection of  $\M$ onto $\mathcal{C}_{lin}$ is given by $\vec^{-1}(\m + \A' (\A \A')^{-1}(\boldsymbol{e}_1 - \A\m))$. This expression simplifies significantly by close inspection. In fact, it is equivalent to setting $m_{11} = 1$ and for $j \geq 2$ replacing $m_{j1}$, $m_{1j}$, and $m_{jj}$ by their average $(m_{jj} + m_{j1} + m_{1j})/3$. Denote this projection with the operator $\mathcal{P}_{lin}(\cdot)$.

Second, \cite{tanaka2014positive} describe a univariate optimization problem that is almost equivalent to projecting $\M$ onto $\mathcal{C}_{sd}$. The only difference is that their solution set also includes the zero-matrix $\boldsymbol{0}$. Assuming that such a limiting case can be safely handled in the implementation, their approach offers a fast projection onto  $\mathcal{C}_{sd}$ even for a moderately large $N$. To describe this approach, consider the spectral decomposition $\M = \Q \Diag(l_1, \dots, l_{N+1}) \Q'$ and the univariate function 
\begin{align*}
\pi(\mu) &= \sum_{i=1}^{N+1} \left[ \left(\mu-l_i\right)^2_+ + \left(l_i - \kappa\mu\right)^2_+ \right],
\end{align*}
where $\Diag(\x)$ is a diagonal matrix with diagonal $\x$ and $(\cdot)_+$ is the positive part operator. The function $\pi(\mu)$ can be minimized very efficiently by solving a series of smaller convex problems, each with a closed form solution. The result is a binary-search-like procedure described by Algorithm \ref{projSD_algo} in Appendix A.
If $\mu^* = \argmin_{\mu \geq 0} \pi(\mu)$ and 
\begin{align*}
\lambda_j^* &:= \begin{cases}
\mu^* & \text{ if } l_j \leq \mu^*\\
\kappa \mu^* & \text{ if } \kappa \mu^*  \leq l_j\\
l_j & \text{ otherwise}
\end{cases}
\end{align*}
for $j = 1, \dots, N+1$,  then $\Q \Diag(\lambda_1^*, \dots, \lambda_{N+1}^*) \Q$ is the projection of $\M$ onto $\mathcal{C}_{sd}$. Call this projection $\mathcal{P}_{sd}(\cdot : \kappa)$.

 Algorithm \ref{proj_algo} uses these projections to solve (\ref{lse_problem}).  Each iteration projects twice on one set and once on the other set. The general form of the algorithm does not specify which projection should be called twice. Therefore, given that $\mathcal{P}_{sd}(\cdot:\kappa)$ takes longer to run than $\mathcal{P}_{lin}(\cdot)$, it is beneficial to choose to call $\mathcal{P}_{lin}(\cdot)$ twice. The complexity of each iteration is determined largely by the spectral decomposition which is fairly fast for moderately large $N$. Overall time to convergence, of course, depends on the choice of the stopping criterion. Many intuitive criteria are possible. Given that $\bOmega_D \in \mathcal{C}_{lin}$ and $\bOmega_C \in \mathcal{C}_{sd}$, the stopping criterion $\max \{(\bOmega_{D}-\bOmega_{C})_{ij}^2\} < \epsilon$ suggests that the return value is in $\mathcal{C}_{sd}$ and close to $\mathcal{C}_{lin}$ in every direction. Based on our experience, the algorithm converges quite quickly. For instance, our implementation in C++ generally solves (\ref{lse_problem}) for $\epsilon = 10^{-5}$ and $N = 100$  in less than a second on a 1.7 GHz Intel Core i5 computer. This code will be made available online upon publication. For the remainder of the paper, projecting $\SS$ onto the feasible region is denoted with the operator $\mathcal{P}_{LSE}(\SS : \kappa)$.

\begin{algorithm}[t]
\caption{This procedure projects $h(\boldsymbol{S})$ onto the intersection $\mathcal{C}_{sd} \cap \mathcal{C}_{lin}$. Denote the projection with $\mathcal{P}_{LSE}(\SS : \kappa)$. Throughout the paper, the stopping criterion is fixed at $\epsilon = 10^{-5}$.}
\label{proj_algo}
\begin{algorithmic}[1]
\Require Unconstrained covariance matrix estimator $\SS$, stopping criterion $\epsilon > 0$, and an upper bound on the condition number $\kappa \geq 1$. 
\Procedure{Directional Alternating Projection Algorithm}{}
\State $\bOmega_{A} \leftarrow  h(\boldsymbol{S})$
\Repeat

\State $\bOmega_{B} \hspace{0.5em} \leftarrow \mathcal{P}_{lin}(\bOmega_{A})$
\State $\bOmega_{C} \hspace{0.5em} \leftarrow \mathcal{P}_{sd}(\bOmega_{B}: \kappa)$
\State $\bOmega_{D} \hspace{0.5em} \leftarrow \mathcal{P}_{lin}(\bOmega_{C})$
\State $\Delta \hspace{1.2em}\leftarrow ||\bOmega_{B} - \bOmega_{C}||_F^2 / \Tr\left[(\bOmega_{B}-\bOmega_{D})' (\bOmega_{B}-\bOmega_{C} )\right]$
\State $\bOmega_{A} \hspace{0.6em} \leftarrow \bOmega_{B} + \Delta(\bOmega_{D}-\bOmega_{B})$
\Until{$\max \left\{\left(\bOmega_{D}-\bOmega_{C}\right)_{ij}^2\right\} < \epsilon$}
\State
\Return $h^{-1}(\bOmega_{C})$
\EndProcedure
\end{algorithmic}
\label{algo}
\end{algorithm}

\subsection{Selecting $\kappa$}
\label{condSelection}
The estimation procedure described in the previous section has one tuning parameter, namely the condition number threshold $\kappa$. This subsection discusses an in-sample approach, called  \textit{conditional validation},
 that can be used for choosing any tuning parameter, such as $\kappa$, under the partial information framework. To motivate, recall that the revealed aggregator $X''$ uses $\bSigma$ to regress  $Z_0$ on the rest of the $Z_{j}$s. 
 Of course, the accuracy of this prediction cannot be known until the actual outcome is observed. However, apart from being unobserved, the variable $Z_0$ is theoretically no different to the other $Z_{j}$s. This suggests the following algorithm: for some value $\nu$ compute $\mathcal{P}_{LSE}(\SS : \nu)$, let each of the $Z_{j}$s in turn play the role of $Z_0$, predict its value based on $Z_{i}$ for $i \neq j$, and choose the value of $\nu$ that yields the best overall accuracy. Even though many accuracy measures could be chosen, this paper uses the conditional log-likelihood. Therefore, if $\Z_j^* = (Z_{j1}, \dots, Z_{jK})'$ collects the $j$th forecaster's information about the $K$ events, the chosen value of $\kappa$ is
  \begin{align}
\kappa_{cov} = \argmax_{\nu \geq 1} \sum_{j=1}^N \ell \left( \Z_{j}^*, \mathcal{P}_{LSE}(\SS : \nu) \big| \Z_i^* \text{ for } i \neq j\right), \label{kappa_cov}
\end{align}
where the log-likelihood is now conditional on $\Z_i^*$s for $i \neq j$ and $\SS$ is computed  based on all the forecasts $\Z_1^*, \dots, \Z_N^*$. Plugging this into the projection algorithm gives the final estimate $\bSigma_{cov} := \mathcal{P}_{LSE}(\SS : \kappa_{cov})$.  

Unfortunately, the optimization problem (\ref{kappa_cov}) is non-convex in $\nu$.  However, as was mentioned before, Algorithm \ref{proj_algo} is fast for moderately sized $N$. Therefore $\kappa$ can be chosen efficiently (possibly in parallel on multicore machines) over a  grid of candidate values. Overall, the idea in conditional validation is similar to cross-validation but, instead of predicting across rows (observations), the prediction is  performed across columns (variables).  This not only mimics the actual process of revealed aggregation but is also likely to be more appropriate for prediction polling that typically involves a large number of forecasters (large $N$) predicting relatively few events (small $K$). 
Furthermore, it has no tuning parameters and remains more stable when $K$ is small; see Appendix B for an illustration of this result under synthetic data. 

\section{APPLICATIONS}
This section applies the partial information framework to different types of real world forecasts. For each type there may be different ways to adopt the Gaussian model. The main point, however, is not to find  the optimal way to do this but rather to give illustrative examples on using the framework and also to show how the resulting partial information aggregators outperform the commonly used measurement error aggregators. 

\label{applications}
\subsection{Probability Forecasts of Binary Outcomes}
\label{binaryReal}
\subsubsection{Dataset}
During the second year of the Good Judgment Project (GJP) the forecasters made probability estimates for $78$ events, each with two possible outcomes. One of these events was illustrated in Figure \ref{Example_prob}. 
Each prediction problem had a timeframe, defined as the number of days between the first day of forecasting and the anticipated resolution day. 
These timeframes varied largely among problems, ranging from 12 days to 519 days with a mean of 185.4 days.
%
During each timeframe the forecasters were allowed to update their predictions as frequently as they liked.  The forecasters knew that their estimates would be assessed for accuracy using the quadratic loss (often  known as the Brier score; see \citealt{brier} for more details). 
This is a proper loss function that incentivized the forecasters to report their true beliefs instead of attempting to game the system. In addition to receiving \$150 for meeting minimum participation requirements that did not depend on prediction accuracy, the forecasters received status rewards for their performance via leader-boards displaying the losses for the best $20$ forecasters. Depending on the details of the reward structure, such a competition for rank may eliminate the truth-revelation property of proper loss functions (see, e.g., \citealt{lichtendahl2007probability}).



This data collection raises several issues. First, given that the current paper does not focus on modeling dynamic data, only forecasts made within some common time interval should be considered.  
 Second, not all forecasters made predictions for all the events. Furthermore, the forecasters generally updated their forecasts infrequently, resulting into a very sparse dataset. Such high sparsity can cause problems in  computing the initial unconstrained estimator $\SS$. 
Evaluating different techniques to handle missing values, however, is well outside the scope of this paper. Therefore, to somewhat alleviate the effect of missing values, only the hundred most active forecasters are considered. This makes sufficient overlap highly likely but, unfortunately, still not guaranteed.

All these considerations lead to a parallel analysis of three scenarios: High Uncertainty (HU), Medium Uncertainty (MU), and Low Uncertainty (LU). Important differences are summarized in Table \ref{scenarios}. Each scenario considers the forecasters' most recent prediction within a different time interval.  For instance, LU only includes each forecaster's most recent forecast during $30-60$ days before the anticipated resolution day. The resulting dataset has $60$ events of which $13$ occurred. In the corresponding $60 \times 100$ table of forecasts, around 42 \% of the values are missing. The other two scenarios are defined similarly.

\begin{table}[h!]
   \centering
   \caption{Summary of the three time intervals analyzed. Each scenario considers the forecasters' most recent forecasts within the given time interval. The value in the parentheses represent the number of events occurred. The final column shows the percent of missing forecasts. }
   \begin{tabular}{l ccc } 
   \hline\hline
      Scenario      & Time Interval & \# of Events & Missing (\%)  \\\hline
      High Uncertainty (HU)        & $90-120$     & $49$ $(10)$ & $51$\\ 
      Medium Uncertainty (MU)       & $60-90$  &  $56$ $(14)$  & $46$ \\
      Low Uncertainty (LU) &  $30-60$ & $60$ $(13)$ & $42$\\ \hline
   \end{tabular}
   \label{scenarios}
\end{table}

\subsubsection{Model Specification and Aggregation}
The first step is to pick a link function and derive a Gaussian model for probability forecasts of binary events. Overall, this construction resembles in many ways the latent variable version of a standard probit model. 
\begin{quote}
\textbf{Model Instance.} Identify the $k$th event with $Y_k \in \{0,1\}$. These outcomes link to the information variables via the following function:
\begin{align*}
Y_k &= g(Z_{0k}) = \begin{cases}
1 & \text{ if } Z_{0k} > t_k\\
0 & \text{ otherwise},\\
\end{cases}
\end{align*}
where $t_k \in \R$ is some threshold value. Therefore the link function $g(\cdot)$ is simply the indicator function $\one_{A_k}$ of the event $A_k = \{Z_{0k} > t_k\}$. This threshold is defined by the prior probability of the $k$th event $\P(Y_k = 1) = \Phi(-t_k)$, where $\Phi(\cdot)$ is the CDF of a standard Gaussian distribution. Given that the thresholds are allowed to vary among the events, each event has its own prior. 
The corresponding probability forecasts $X_{jk} \in [0,1]$ are
\begin{align*}
X_{jk} &= \E(Y_k | Z_{jk}) = \Phi\left( \frac{Z_{jk} - t_k}{\sqrt{1-\delta_j}}\right).
\end{align*}
In a similar manner, 
 the revealed aggregator $X_k'' \in [0,1]$ for event $k$ is
\begin{align}
X_{k}'' &= \E(Y_k | \Z_k) = \Phi\left( \frac{\diag(\bSigma)'\bSigma^{-1}\Z_k - t_k}{\sqrt{1-\diag(\bSigma)'\bSigma^{-1}\diag(\bSigma)}}\right). \label{RevAgg}
\end{align}
\end{quote}




All the parameters of this model can be estimated from the data. The first step is to specify a version of the  unconstrained estimate $\SS$. If the $t_k$'s do not change much, a reasonable and simple estimate is obtained  by transforming the sample covariance matrix  $\SS_P$ of the probit scores $P_{jk} := \Phi^{-1}(X_{jk})$. More specifically,  if $\D := \Diag( \dd ) \Diag(\one + \dd)^{-1}$, where $\dd = \diag(\SS_P)$, then an unconstrained estimator of $\bSigma$ is given by $\SS = (\I_N-\D)^{1/2} \SS_P (\I_N-\D)^{1/2}$. Recall that the GJP data holds many missing values. This is handled by estimating each pairwise covariance in $\SS_P$ based on all the events for which both forecasters made predictions. 
Next, compute $\bSigma_{cov}$, where $\kappa_{cov}$ is chosen over a grid of $100$ candidate values between $10$ and $1,000$. 
Finally, the threshold $t_k$ can be estimated by letting $\PP_k = (P_{1k}, \dots, P_{Nk})'$, observing that $-\Diag(\one-\diag(\bSigma))^{1/2}\PP_{k} \sim \mathcal{N}_N(t_k \one_N, \bSigma)$,
%
and computing the precision-weighted average: 
\begin{align*}
\hat{t}_k &= -\frac{\PP_{k}' \Diag(\one-\diag(\bSigma_{cov}))^{1/2} \bSigma_{cov}^{-1} \one}{\one' \bSigma_{cov}^{-1} \one}.
\end{align*}
 If $\PP_k$ has missing values, the corresponding rows and columns of $\bSigma_{cov}$ are dropped. Intuitively, this estimator gives more weight to the forecasters with very little information. These estimates are then plugged in to (\ref{RevAgg}) to get the revealed aggregator  $X_{cov}''$.

This aggregator is benchmarked against the state-of-the-art measurement-error aggregators, namely the average probability, median probability, average probit-score, and average log-odds. Unequally weighted averages were not considered because it is unclear how the weights would be determined based on forecasts alone, and even if this could be done somehow (perhaps based on self-assessment or organizational status), using unequal weights often leads to no or very small performance gains \citep{rowse1974comparison, ashton1985aggregating, flores1989subjective}. To avoid infinite log-odds
and probit scores, extreme forecasts $X_{jk} = 0$ and $1$ were
censored to $X_{jk} = 0.001$ and $0.999$, respectively. The results remain insensitive to the exact choice of censoring as long as this is done in a reasonable manner to keep the extreme probabilities from becoming highly influential in the logit- or probit-space. The accuracy of the aggregates is measured with the average root-mean-squared-error (RMSE). Note that this is nothing but the square root of the commonly used Brier score. 
 Instead of considering all the forecasts at once, the aggregators are evaluated under different $N$ via repeated subsampling of the $100$ most active forecasters; that is, choose $N$ forecasters uniformly at random, aggregate their forecasts, and compute the RMSE. This is repeated 1,000 times with $N = 5, 10, \dots, 65$ forecasters. Due to high computational cost, the simulation was stopped after $N = 65$. In the rare occasion where no pairwise overlap is available between one or more pairs of the selected forecasters, the subsampling is repeated until all pairs have at least one problem in common. 

\begin{figure}[t!]
\vspace{-2em}
\centering
\hspace*{1em} 	\includegraphics[width=\textwidth]{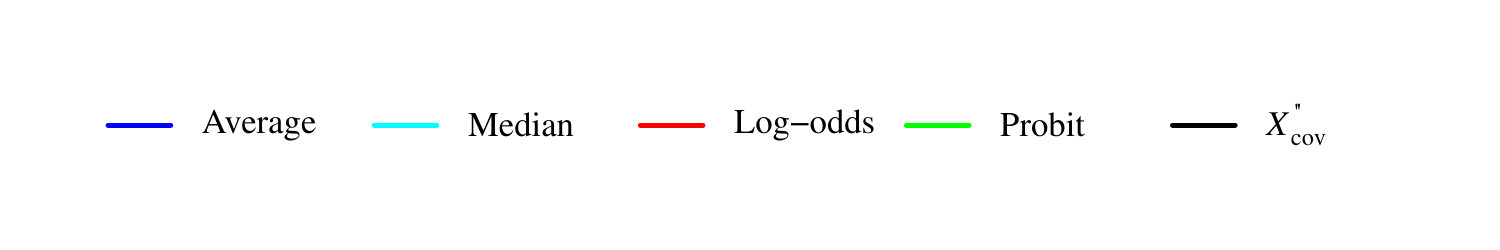} 
\vspace{-4em}

        \centering
        \begin{subfigure}{0.32\textwidth}
                \includegraphics[width=\textwidth]{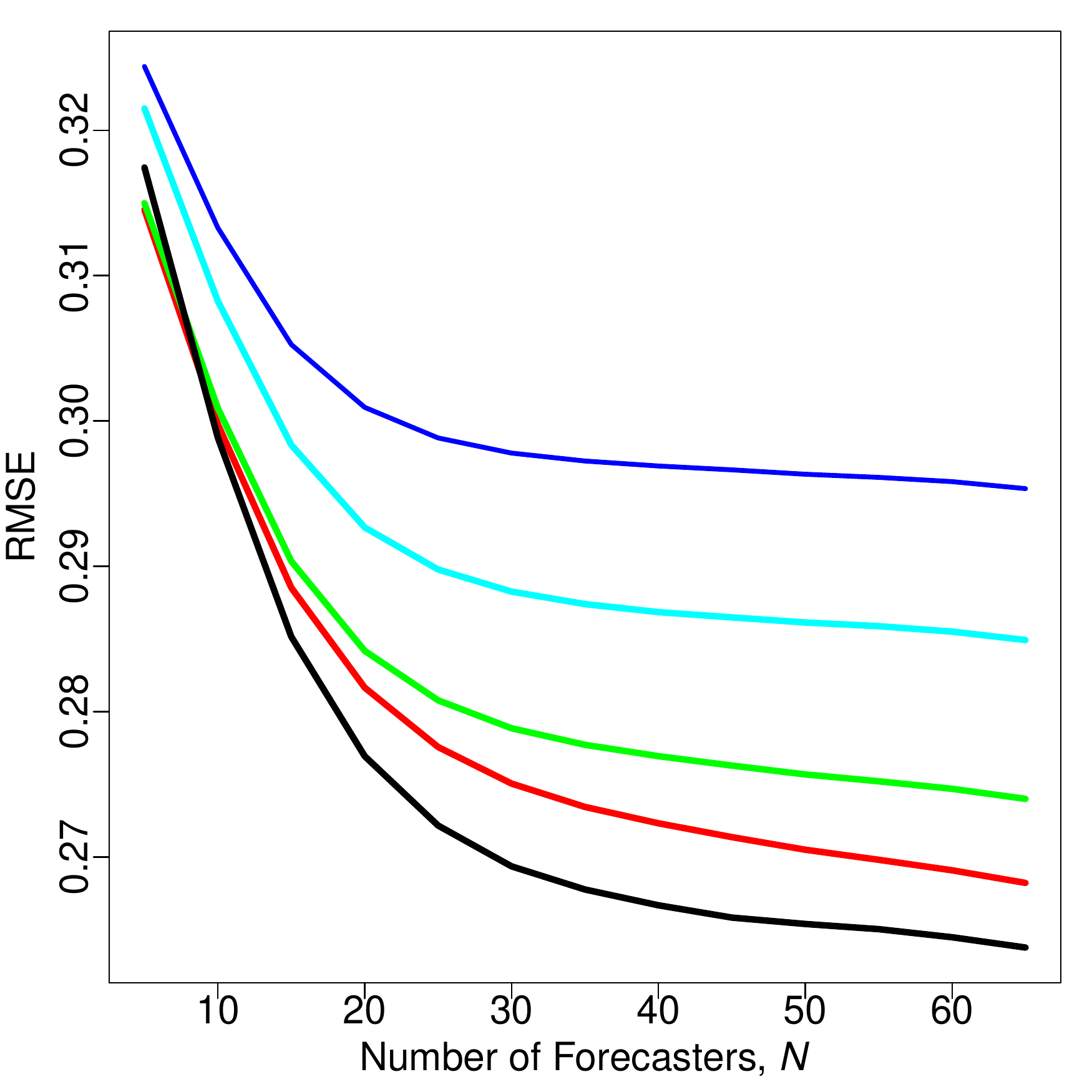}
                \caption{High Uncertainty (HU)}
                                \label{highUnc}
        \end{subfigure}
        \begin{subfigure}{0.32\textwidth}
                \includegraphics[width=\textwidth]{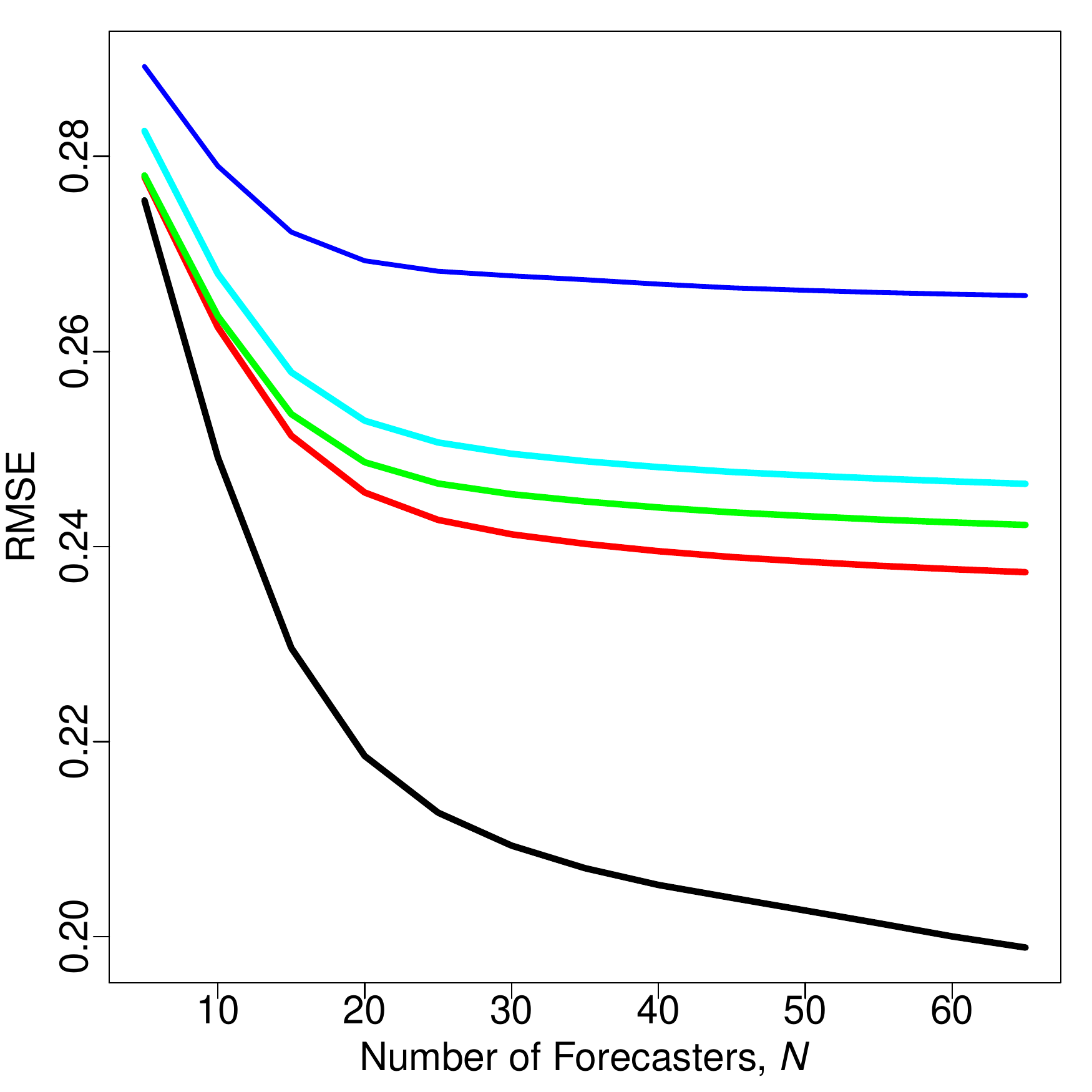}
                \caption{Medium Uncertainty (MU)}
                                \label{medUnc}
        \end{subfigure}
        \begin{subfigure}{0.32\textwidth}
                \includegraphics[width=\textwidth]{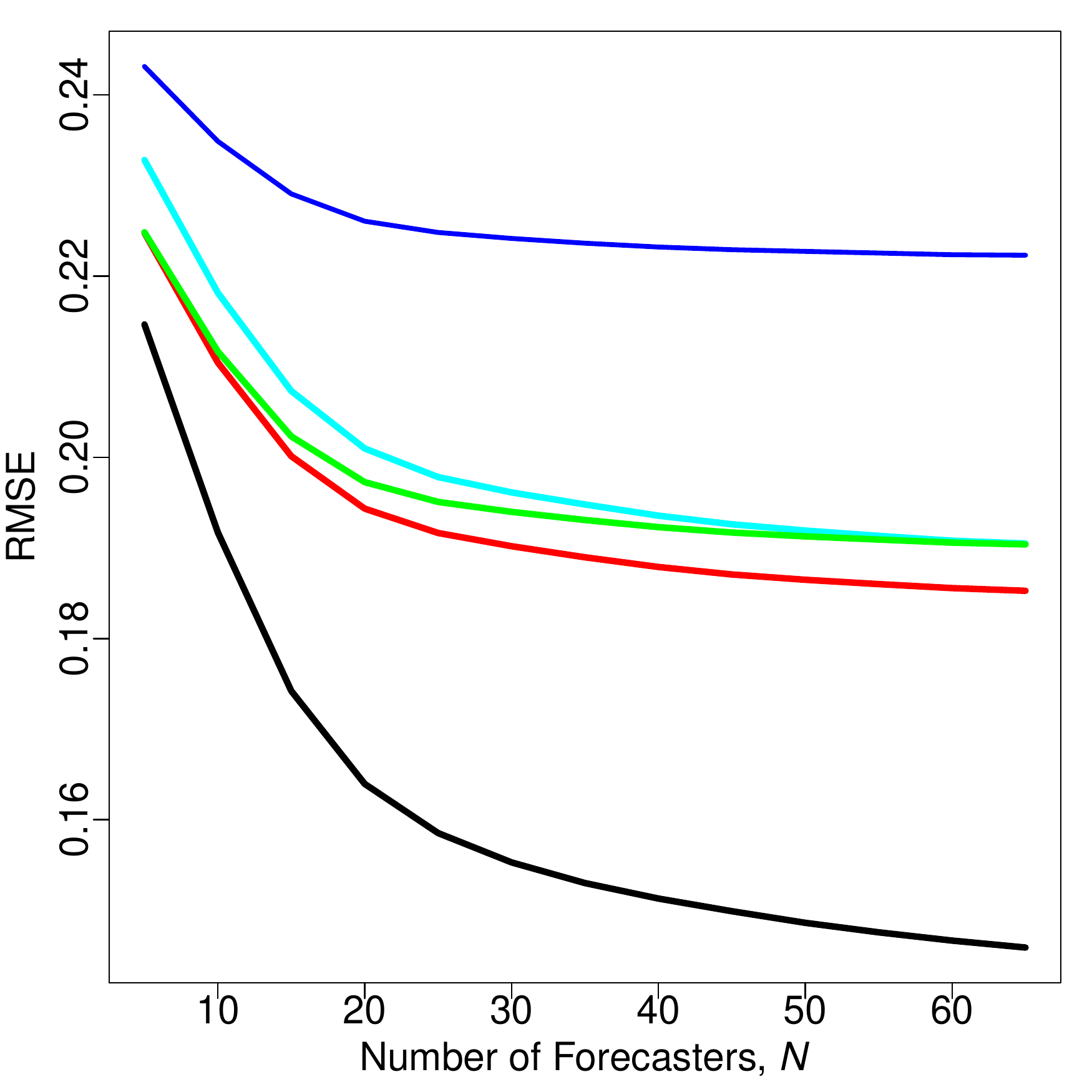}
                \caption{Low Uncertainty (LU)}
                                \label{lowUnc}
        \end{subfigure}             
        
        \caption{Average prediction accuracy over the 1,000 sub-samplings of the forecasters. See Table \ref{scenarios} for descriptions of the different scenarios.}
        \label{simReal}
\end{figure}


Figure \ref{simReal} shows the average RMSEs under the three scenarios described in Table \ref{scenarios}.  Here a reasonable upper bound is given by $0.5$ as this is the RMSE one would receive by constantly predicting $0.5$. All presented scores, however, are well below it and improve uniformly from left to right, that is, from HU to LU. This reflects the decreasing level of uncertainty. 
In all the figures the measurement-error aggregators rank in the typical order (from worst to best): average probability, median probability, average probit, and average log-odds. Regardless of the level of uncertainty, the revealed aggregator $X_{cov}''$ outperforms the averaging aggregators as long as $K \geq 10$. 
The relative advantage, however, increases from HU to LU. More specifically, the improvement from Log-odds to $X_{cov}''$ is about $2$\%, $17$\%, and $21$\% in HU, MU, and LU, respectively. This trend can be explained by several reasons. First, as can be seen in Table \ref{scenarios}, the amount of data increases from HU to LU. This yields a better estimate of $\bSigma$ and hence more accurate revealed aggregation.
Second, the forecasters are more likely to be well-calibrated under MU and LU than under HU (see, e.g., \citealt{braun1992case}).
 Third, under HU the events are still inherently very uncertain. Consequently, the forecasters are unlikely to hold much useful information as a group. Under such low information diversity, measurement-error aggregators generally perform relatively well (\citealt{satopaamodeling}).  In the contrary, under MU  the events have lost a part of their inherent uncertainty, allowing some forecasters to possess useful private information. These individuals are then prioritized by $X_{cov}''$ while the averaging-aggregators continue treating all forecasts equally. Consequently, the performance of the measurement error aggregators plateaus after $N = 30$ or so.  Therefore having more than about $30$ forecasters does not make a difference if one is determined to aggregate their predictions using the measurement error techniques; a similar results was reported by \citealt{satopaa}. In contrast, however, the RMSE of $X_{cov}''$ continues to improve linearly in $N$, suggesting that $X_{cov}''$ is able to find some residual information in each additional forecaster and use this to increase its performance advantage.

\subsubsection{Information Diversity}

The GJP assigned the forecasters to make predictions either in isolation or in teams. Furthermore, after the first year of the tournament, the top 2\% forecasters were elected to the elite group of ``super-forecasters.'' These super-forecasters then worked in exclusive teams to make highly accurate predictions on the same events as the rest of the forecasters. Overall, these assignments directly suggest a level of information overlap. In particular, recalling the interpretation of $\bSigma$ from Section \ref{context}, super-forecasters can be expected to have the highest $\delta_j$s and forecasters in the same team should have a relatively high $\rho_{ij}$. This subsection examines how well $\bSigma_{cov}$ aligns with this prior knowledge about the forecasters' information structure. 


\begin{figure}[t!]
        \centering
        \begin{subfigure}[b]{0.51\textwidth}
                \includegraphics[width=\textwidth]{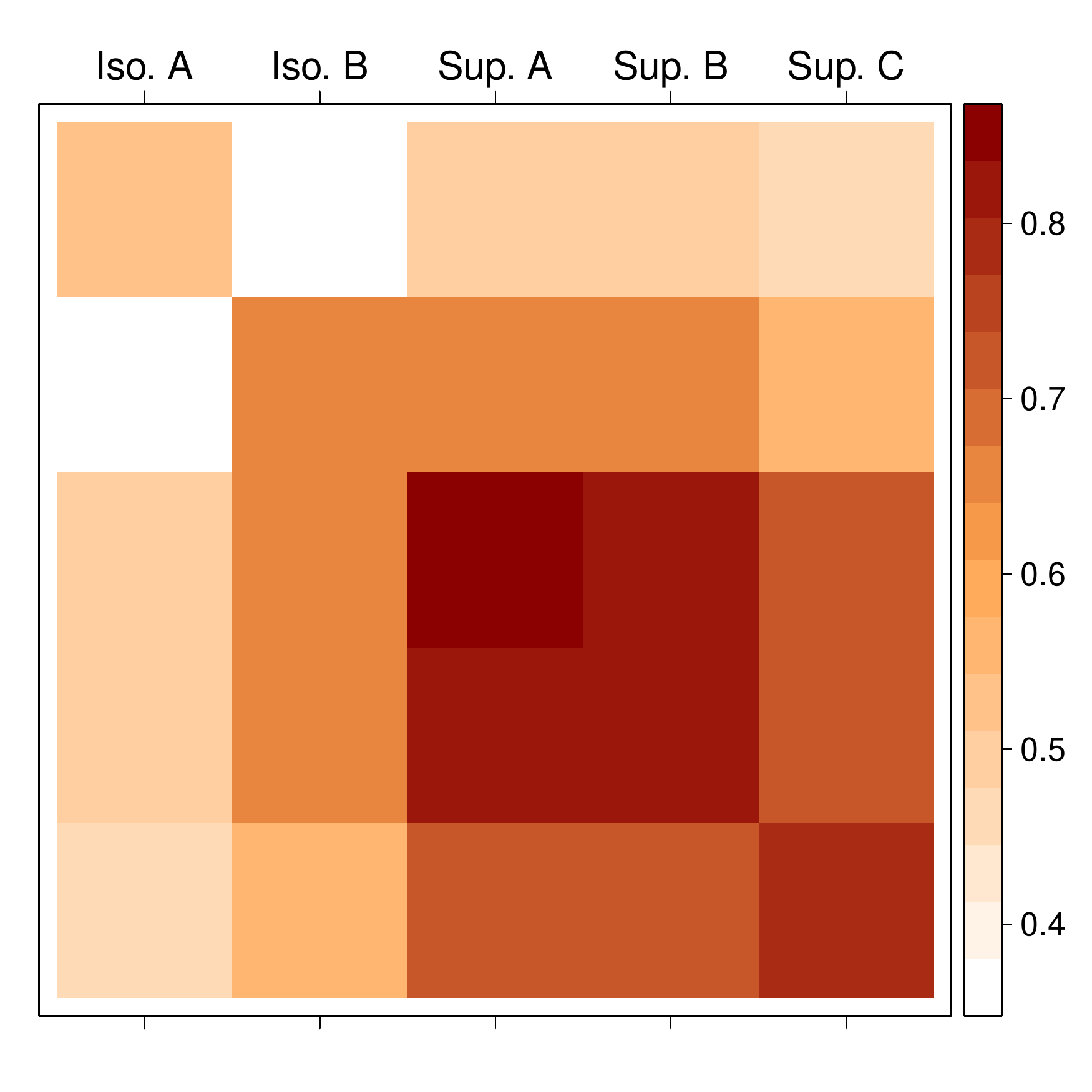}
                \caption{$\bSigma_{cov}$ for the five most active forecasters}
                \label{top5}
        \end{subfigure}%
        \begin{subfigure}[b]{0.49\textwidth}
        \hspace*{-11em} 	\includegraphics{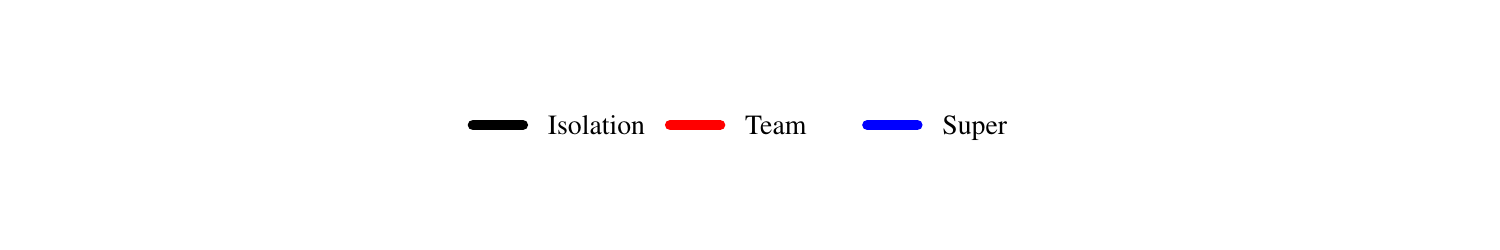} 
\vspace{-4.5em}

                \includegraphics[width=\textwidth]{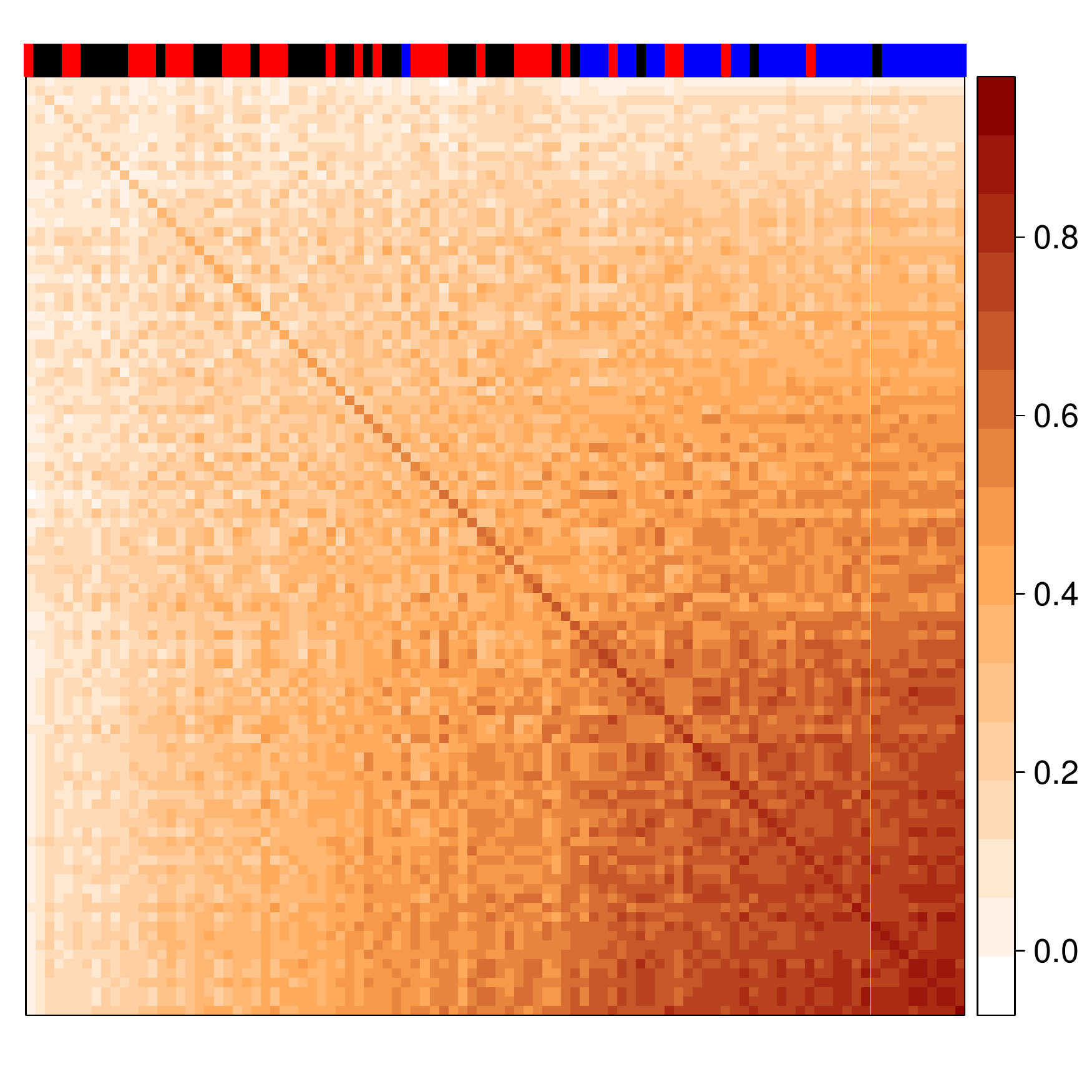}
            \caption{$\bSigma_{cov}$ for all 100 forecasters shows high information diversity.}
                                \label{top100}
	\vspace{-1.1em}
       \end{subfigure}
      \vspace{1 em}
        \caption{The estimated information structure $\bSigma$ under the LU scenario. Each forecaster worked either in isolation, in a non-super-forecaster team, or in a super-forecaster team. The super-forecasters generally have more information than the forecasters working in isolation.}
        \label{SynthPlot}
\end{figure}


For the sake of brevity, only the LU scenario is analyzed as this is where $X_{cov}''$ presented the highest relative improvement. 
The associated 100 forecasters involve 36 individuals predicting in isolation, 33 forecasting team-members (across 24 teams), and 31 super-forecasters (across 5 teams).  Figure \ref{top5} displays $\bSigma_{cov}$ for the five most active forecasters. This group involves two forecasters working in isolation (Iso. A and B) and three super-forecasters (Sup. A, B, and C), of whom the super-forecasters A and B are in the same team. Overall, $\bSigma_{cov}$ agrees with this classification: the only two team members, namely Sup. A and B have a relatively high information overlap. In addition, the three super-forecasters are more informed than the non-super-forecasters. Such a high level of information unavoidably leads to higher information overlap with the rest of the forecasters. 

By and large, this agreement generalizes to the entire group of forecasters. To illustrate, Figure  \ref{top100} displays $\bSigma_{cov}$ for all the 100 forecasters. The information structure has been ordered with respect to the diagonal such that the more informed forecasters appear on the right. Furthermore, a colored rug has been appended on the top. This rug shows whether each forecaster worked in isolation, in a non-super-forecaster team, or in a super-forecaster team. Observe that the super-forecasters are mostly situated on the right among the most informed forecasters. The average estimated $\delta_j$ among the super-forecaster is $0.80$.  On the other hand, the average estimated $\delta_j$ among the individuals working in isolation or in non-super-forecaster teams are $0.47$ and $0.50$, respectively. Therefore working in a team makes the forecasters' predictions, on average, slightly more informed. 

In general, a plot such as Figure  \ref{top100} is useful for assessing the level of information diversity among the forecasters: the further away it is from a monochromatic plot, the higher the information diversity. That being said, the colorful Figure  \ref{top100} suggests that the GJP forecasters have high information diversity. This makes sense as these forecasters were asked to make predictions about international political events. Given that on such events the forecasters' background knowledge, education, how closely they follow the news, and so on matter, one should expect a high level of information diversity. Therefore not only does $X_{cov}''$ clearly outperform the common measurement error aggregators in terms of prediction accuracy but the Gaussian model also captures true structure in the data. 


\subsection{Point Forecasts of Continuous Outcomes}
\label{continuousReal}
\subsubsection{Dataset}
\label{continuousRealData}
\citet{moore2008use} hired $415$ undergraduates from Carnegie Mellon University to guess the weights of $20$ people based on a series of pictures. These forecasts were illustrated in Figure \ref{Example_point}. The target people were between 7 and 62 years old and had weights ranging from $61$ to $230$ pounds, with a mean of $157.6$ pounds. All the students were shown the same pictures and hence given the exact same information. 
Therefore any information diversity arises purely from the participants' decisions to use different subsets of the same information. 
Consequently, information diversity is likely to be low compared to Section \ref{binaryReal} where diversity also stemmed from differences in the information available to the forecasters.


Unlike in Section  \ref{binaryReal}, the Gaussian model can be applied almost directly to the data. 
Only the effect of extreme values was reduced via a $90$\% Winsorization \citep{hastings1947low}. This handled some obvious outliers. For instance, the original dataset contained a few estimates above $1000$ pounds and as low as $10$ pounds.  Winsorization generally improved the performance of all the competing aggregators.



\subsubsection{Model Specification and Aggregation}
\begin{samepage}
\begin{quote}
\textbf{Model Instance.} Suppose $Y_k$ and $X_{jk}$ are real-valued. If the proper non-informative prior distribution of $Y_k$ is $ \mathcal{N}(\mu_{0k}, \sigma_0^2)$, then $Y_k = g(Z_{0k}) = Z_{0k}\sigma_0 + \mu_{0k}$. Consequently, $X_{jk} =  \E(Y | Z_{jk}) = Z_{jk}\sigma_0 + \mu_{0k}$ for all $j = 1, \dots, N$. Therefore $X_j \sim \mathcal{N}(\mu_{0k}, \sigma_j^2)$ for some $\sigma_j^2 \leq \sigma_0^2$.  
%
  If $\Z_k = \left(Z_{1k}, \dots, Z_{Nk}\right)'$, then the revealed aggregator for the $k$th event is 
\begin{align}
X_k'' =  \E\left(Y_k | \Z_k\right) &= \diag(\bSigma)' \bSigma^{-1} \Z_k \sigma_0 + \mu_{0k}. \label{revPoint}
\end{align}
\end{quote}
\end{samepage}

Under this model the prior distribution of $Y_k$ is specified by $\mu_{0k}$ and $\sigma_0^2$. Given that $\E(X_{jk}) = \mu_{0k}$ for all $j = 1, \dots, N$, the sample average $\hat{\mu}_{0k} = \sum_{j=1}^N X_{jk}/N$ provides an initial estimate of $\mu_{0k}$. 
The value of $\sigma_0^2$ can be estimated by assuming a distribution for the $\sigma_j^2$s. More specifically, let $\sigma_j^2$ be i.i.d. on the interval $[0,\sigma_0^2]$ and use the resulting likelihood to estimate $\sigma_0^2$. For instance, a non-informative choice is to assume $\sigma_j^2 \stackrel{i.i.d.}{\sim} \mathcal{U}(0,\sigma_0^2)$, which leads to the maximum likelihood estimator $\max\{\sigma_j^2\}$. This has a downward bias that can be corrected by a multiplicative factor of $(N+1)/N$. Therefore, replacing $\sigma_j^2$ with the sample variance  $s_j = \sum_{k=1}^K (X_{jk}-\hat{\mu}_{0k})^2/(K-1)$ gives the final estimate $\hat{\sigma}_0^2 = (N+1)/N\max\{s_j\}$. Using these estimates, the $X_{jk}$s can be transformed into the $Z_{jk}$s whose sample covariance matrix $\SS_Z$ provides the unconstrained estimator for the projection algorithm.  The value of $\kappa_{cov}$ is chosen over a grid of $10$ values between $10$ and $10,000$. Once $\bSigma_{cov}$ has been computed, the prior means are updated with the precision-weighted averages $\hat{\mu}_{0k} = (\X_{k}' \bSigma_{cov}^{-1} \one_N)/(\one_N' \bSigma_{cov}^{-1} \one_N)$. In the end, all these estimates are plugged in (\ref{revPoint}) to get the revealed aggregator $X_{cov}''$.


\begin{figure}[t!]
\captionsetup{width=0.48\textwidth}
\vspace{-2em}
        \centering
        \begin{minipage}[b]{0.495\textwidth}
	\hspace*{1em} 	\includegraphics[width=\textwidth]{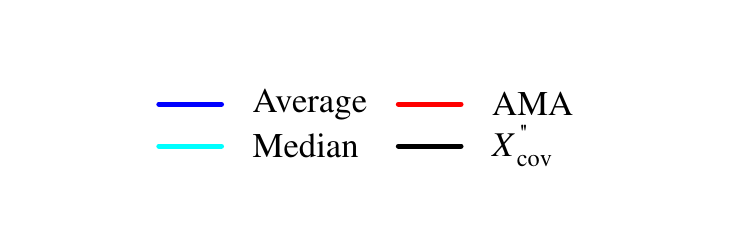} 
	\vspace{-4.5em}

                \includegraphics[width=\textwidth]{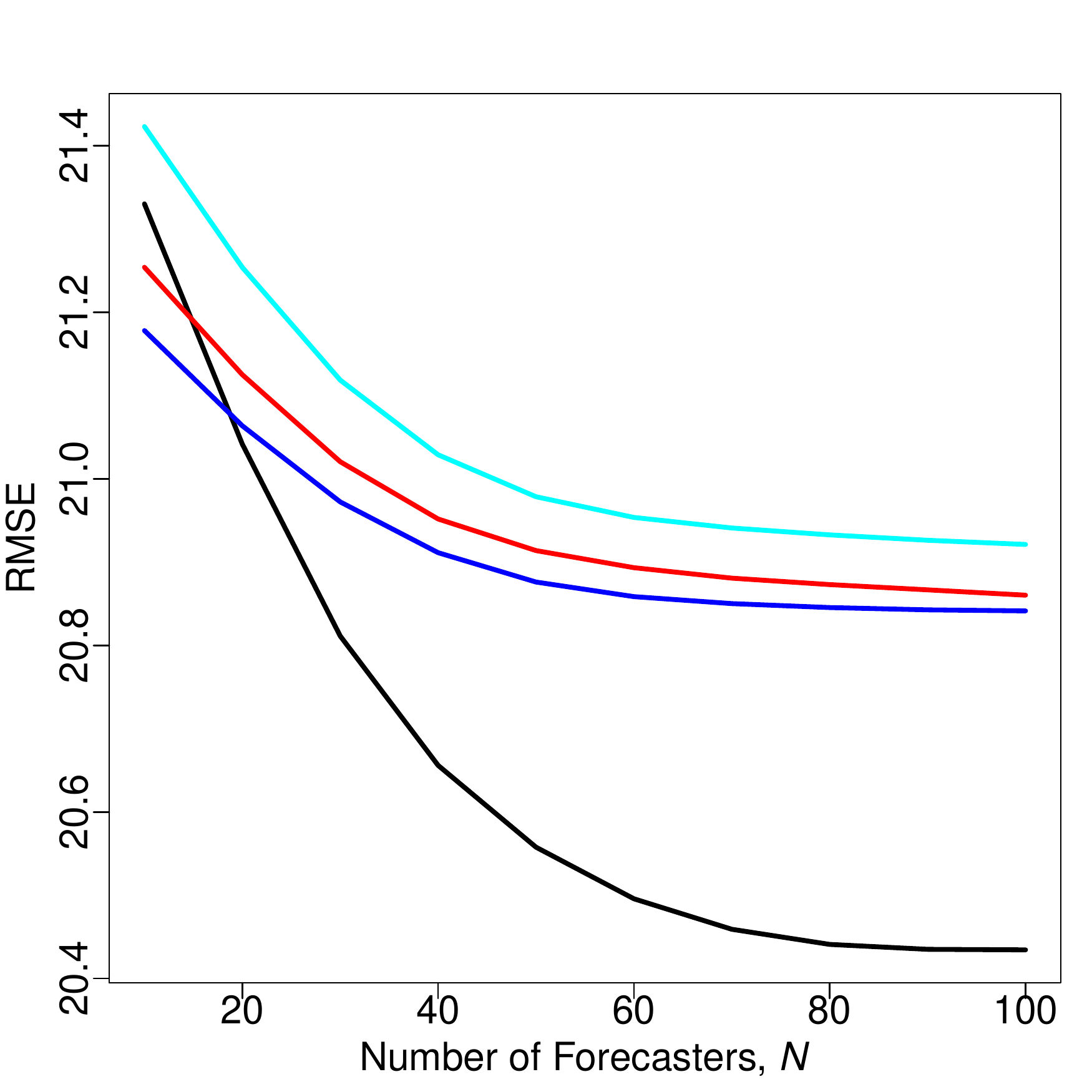}
                \caption{Average prediction accuracy}
                \label{pointsAcc}
        \end{minipage}%
        \begin{minipage}[b!]{0.495\textwidth}
  	\vspace{-19em}
               \includegraphics[width=\textwidth]{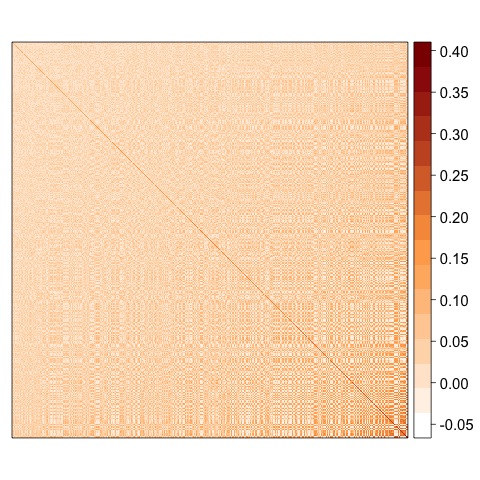}
            \caption{$\bSigma_{cov}$ for all 416 forecasters shows low information diversity.}
                                \label{ID416}
      \end{minipage}
\end{figure}

This aggregator is compared against the average, median, and average of the median and average (AMA). The last competitor, namely AMA is a heuristic aggregator that \citet{lobo2010human} showed to work particularly well on many different real-world forecasting datasets. In this section the overall accuracy is measured with the RMSE averaged over $10,000$ sub-samplings of the $416$ participants. That is, each iteration chooses $N$ participants uniformly at random, aggregates their forecasts, and computes the RMSE. The size of the sub-samples is varied between $10$ and $100$ with increments of $10$. These scores are presented in Figure \ref{pointsAcc}. The average outperforms the median across all $N$. The performance of AMA falls between that of average and median, reflecting its nature as a compromise of the two. The revealed aggregator $X_{cov}''$ is the most accurate once $N > 10$. The relatively worse performance at $N = 10$ suggests that $10$ observations is not enough to estimate $\hat{\mu}_{0k}$ accurately.
As $N$ approaches $100$, however, $X_{cov}''$ collects information efficiently and increases the performance advantage against the other aggregators. 


 Figure \ref{ID416} shows $\bSigma_{cov}$ for all the 416 forecasters. Similarly to before, the matrix has been ordered such that the most knowledgeable forecasters are on the right. Overall, this plot is much more monochromatic than the one presented earlier in Figure \ref{top100}, suggesting that information diversity among the 416 students is rather lower.  This aligns with the expectations laid out earlier in Section \ref{continuousRealData}. If there were no information diversity, i.e., all the forecasters used the same information, then averaging aggregators, such as the simple average, would perform very well \citep{satopaamodeling}. Such a limiting case, however, is rarely encountered in practice. Often at least some information diversity is present. The results in the current section show that the revealed aggregator does not require extremely high information diversity in order to outperform the measurement-error aggregators. 
 


\section{DISCUSSION}
\label{discussion}
This paper introduced the partial information framework for modeling forecasts from different types of  prediction polls. Even though the framework can be used for theoretical analysis and studying information among groups of experts, the main focus was on model-based aggregation of forecasts. Such aggregators do not require a training set. Instead, they operate under a model of forecast heterogeneity and hence can be applied to forecasts alone. Under the partial information framework, all forecast heterogeneity stems from differences in the way the forecasters use information. Intuitively, this is more plausible at the micro-level than the historical measurement error. To facilitate practical applications, the partial information framework motivates and describes the forecasters' information with a patterned covariance matrix (Equation \ref{cov_str}). A correctional procedure was proposed (Algorithm \ref{proj_algo}) as a general tool for estimating these information structures. This procedure inputs any covariance estimator and modifies it minimally such that the final output represents a physically feasible allocation of information.
 Even though the general partial information framework describes an optimal aggregator, it is generally too abstract to be directly applied in practice. 
%
%
%
%
As a solution, this paper discusses a close yet practical specification within the framework, known as the Gaussian model (Section \ref{gaussian}). 
  The Gaussian model permits a closed-form solution for the optimal aggregator and extends to  different types of forecast-outcome pairs via a link function. These partial information aggregators were evaluated against the common measurement error aggregators on two different real-world  (Section \ref{applications}) prediction polls. 
  In each case the Gaussian model outperformed the typical measurement-error-based aggregators, suggesting that information diversity is more important for modeling forecast heterogeneity. 

%
%
%

Generally speaking, partial information aggregation works well because it downweights pairs or sets of forecasters that share more information and
      upweights ones that have unique information (or choose to attend to unique information as is the case, e.g., in Section \ref{continuousReal}, where forecasters made judgments based on the same pictures). This is very different from measurement-error aggregators that assume all forecasters to have the same information and hence consider them equally important. While simple measurement-error techniques, such as the average or median, can work well when the forecasters truly operate on the same information set, in real-world prediction polls participants are more likely to have unequal skill and information sets. Therefore prioritizing is almost certainly called for. Of course, the more diverse these sets are, the better the partial information aggregators can be expected to perform relative to the measurement error aggregators. To illustrate this result, compare the relative performances in Section \ref{binaryReal} (high information diversity) against those in Section \ref{continuousReal} (low information diversity).
      
	Overall, the partial information framework can be applied and extended in many different ways. For instance, in this paper the $j$th forecaster's prediction was assumed to be the expectation of $Y$ after observing some partial information $\F_j$. In some applications, however, other constructs, such as the conditional median or other quantiles, may be more appropriate. Such extensions can be handled by considering the distribution of $Y | \F_j$ and then equating the $j$th forecaster's prediction to any desired functional of this distribution. This is particularly easy under the Gaussian model, where  $Y | \F_j$ conveniently follows a Gaussian distribution. 	

      In terms of future research, the partial information framework offers both theoretical and empirical directions. One theoretical avenue involves estimation of information overlap. In some cases the higher order overlaps have been found to be irrelevant to aggregation. For instance, \cite{degroot1991optimal} show that the pairwise conditional (on the truth) distributions of the forecasts are sufficient for computing the optimal weights of a weighted average. Theoretical results on the significance or insignificance of higher order overlaps under the partial information framework would be desirable. Given that the Gaussian model can only accommodate pairwise information overlap, such a result would reveal the need of a specification that is more complex than the Gaussian model.  
      

A promising empirical direction is the Bayesian approach.  These techniques are very natural for fitting hierarchical models such as the ones discussed in this paper.  Furthermore, in many applications with small or moderately sized datasets, Bayesian methods have been found to be more stable than the likelihood-based alternatives. Therefore, given that the number of forecasts in a prediction poll is typically quite small, a Bayesian approach is likely to improve the quality of the final aggregate. This would involve  developing a prior distribution for the information structure -- a problem that seems interesting in itself. 
%
Overall, this avenue should certainly be pursued, and the results tested against other high performing aggregators.
        

\bibliographystyle{apalike}
\bibliography{biblio}		

\begin{thebibliography}{}

\bibitem[Ashton and Ashton, 1985]{ashton1985aggregating}
Ashton, A.~H. and Ashton, R.~H. (1985).
\newblock Aggregating subjective forecasts: Some empirical results.
\newblock {\em Management Science}, 31(12):1499--1508.

\bibitem[Banerjee et~al., 2005]{banerjee2005optimality}
Banerjee, A., Guo, X., and Wang, H. (2005).
\newblock On the optimality of conditional expectation as a bregman predictor.
\newblock {\em Information Theory, IEEE Transactions on}, 51(7):2664--2669.

\bibitem[Braun and Yaniv, 1992]{braun1992case}
Braun, P.~A. and Yaniv, I. (1992).
\newblock A case study of expert judgment: Economists' probabilities versus
  base-rate model forecasts.
\newblock {\em Journal of Behavioral Decision Making}, 5(3):217--231.

\bibitem[Breiman, 1996]{breiman1996stacked}
Breiman, L. (1996).
\newblock Stacked regressions.
\newblock {\em Machine Learning}, 24(1):49--64.

\bibitem[Brier, 1950]{brier}
Brier, G.~W. (1950).
\newblock Verification of forecasts expressed in terms of probability.
\newblock {\em Monthly Weather Review}, 78:1--3.

\bibitem[Broecker, 2012]{jolliffe2012forecast}
Broecker, J. (2012).
\newblock {\em Forecast verification: a practitioner's guide in atmospheric
  science}, chapter 7.2.2, pages 121--122.
\newblock John Wiley \& Sons, Chichester, UK, 2nd edition.

\bibitem[Broomell and Budescu, 2009]{broomell2009experts}
Broomell, S.~B. and Budescu, D.~V. (2009).
\newblock Why are experts correlated? {Decomposing} correlations between
  judges.
\newblock {\em Psychometrika}, 74(3):531--553.

\bibitem[Dawid et~al., 1995]{dawid1995coherent}
Dawid, A., DeGroot, M., Mortera, J., Cooke, R., French, S., Genest, C.,
  Schervish, M., Lindley, D., McConway, K., and Winkler, R. (1995).
\newblock Coherent combination of experts' opinions.
\newblock {\em TEST}, 4(2):263--313.

\bibitem[Dawid, 1982]{dawid1982well}
Dawid, A.~P. (1982).
\newblock The well-calibrated bayesian.
\newblock {\em Journal of the American Statistical Association},
  77(379):605--610.

\bibitem[DeGroot and Mortera, 1991]{degroot1991optimal}
DeGroot, M.~H. and Mortera, J. (1991).
\newblock Optimal linear opinion pools.
\newblock {\em Management Science}, 37(5):546--558.

\bibitem[Di~Bacco et~al., 2003]{Bacco}
Di~Bacco, M., Frederic, P., and Lad, F. (2003).
\newblock Learning from the probability assertions of experts.
\newblock Research Report.
\newblock Available at:
  http://www.math.canterbury.ac.nz/research/ucdms2003n6.pdf.

\bibitem[Dowie, 1976]{dowie1976efficiency}
Dowie, J. (1976).
\newblock On the efficiency and equity of betting markets.
\newblock {\em Economica}, 43(170):139--150.

\bibitem[Flores and White, 1989]{flores1989subjective}
Flores, B.~E. and White, E.~M. (1989).
\newblock Subjective versus objective combining of forecasts: an experiment.
\newblock {\em Journal of Forecasting}, 8(3):331--341.

\bibitem[Foster and Vohra, 1998]{foster1998asymptotic}
Foster, D.~P. and Vohra, R.~V. (1998).
\newblock Asymptotic calibration.
\newblock {\em Biometrika}, 85(2):379--390.

\bibitem[Gigerenzer et~al., 1991]{gigerenzer1991probabilistic}
Gigerenzer, G., Hoffrage, U., and Kleinb{\"o}lting, H. (1991).
\newblock Probabilistic mental models: a brunswikian theory of confidence.
\newblock {\em Psychological Review}, 98(4):506.

\bibitem[Goel et~al., 2010]{goel2010prediction}
Goel, S., Reeves, D.~M., Watts, D.~J., and Pennock, D.~M. (2010).
\newblock Prediction without markets.
\newblock In {\em Proceedings of the 11th ACM conference on Electronic
  commerce}, pages 357--366. ACM.

\bibitem[Gubin et~al., 1967]{gubin1967method}
Gubin, L., Polyak, B., and Raik, E. (1967).
\newblock The method of projections for finding the common point of convex
  sets.
\newblock {\em USSR Computational Mathematics and Mathematical Physics},
  7(6):1--24.

\bibitem[Hastings et~al., 1947]{hastings1947low}
Hastings, C., Mosteller, F., Tukey, J.~W., and Winsor, C.~P. (1947).
\newblock Low moments for small samples: A comparative study of order
  statistics.
\newblock {\em The Annals of Mathematical Statistics}, 18(3):413--426.

\bibitem[Hong and Page, 2009]{hong2009interpreted}
Hong, L. and Page, S. (2009).
\newblock Interpreted and generated signals.
\newblock {\em Journal of Economic Theory}, 144(5):2174--2196.

\bibitem[Hwang, 2004]{hwang2004cauchy}
Hwang, S.-G. (2004).
\newblock Cauchy's interlace theorem for eigenvalues of hermitian matrices.
\newblock {\em American Mathematical Monthly}, 111:157--159.

\bibitem[Juslin, 1993]{juslin1993explanation}
Juslin, P. (1993).
\newblock An explanation of the hard-easy effect in studies of realism of
  confidence in one's general knowledge.
\newblock {\em European Journal of Cognitive Psychology}, 5(1):55--71.

\bibitem[Keren, 1987]{keren1987facing}
Keren, G. (1987).
\newblock Facing uncertainty in the game of bridge: A calibration study.
\newblock {\em Organizational Behavior and Human Decision Processes},
  39(1):98--114.

\bibitem[Keren, 1991]{keren1991calibration}
Keren, G. (1991).
\newblock Calibration and probability judgements: Conceptual and methodological
  issues.
\newblock {\em Acta Psychologica}, 77(3):217--273.

\bibitem[Koriat et~al., 1980]{koriat1980reasons}
Koriat, A., Lichtenstein, S., and Fischhoff, B. (1980).
\newblock Reasons for confidence.
\newblock {\em Journal of Experimental Psychology: Human learning and memory},
  6(2):107.

\bibitem[Kruglanski, 1990]{kruglanski1990motivations}
Kruglanski, A.~W. (1990).
\newblock {\em Motivations for judging and knowing: Implications for causal
  attribution}, volume~2, pages 333--368.
\newblock Guilford Press, New York, NY, US.

\bibitem[Langford et~al., 2001]{langford2001property}
Langford, E., Schwertman, N., and Owens, M. (2001).
\newblock Is the property of being positively correlated transitive?
\newblock {\em The American Statistician}, 55(4):322--325.

\bibitem[Lichtendahl~Jr and Winkler, 2007]{lichtendahl2007probability}
Lichtendahl~Jr, K.~C. and Winkler, R.~L. (2007).
\newblock Probability elicitation, scoring rules, and competition among
  forecasters.
\newblock {\em Management Science}, 53(11):1745--1755.

\bibitem[Lichtenstein and Fischhoff, 1977]{lichtenstein1977those}
Lichtenstein, S. and Fischhoff, B. (1977).
\newblock Do those who know more also know more about how much they know?
\newblock {\em Organizational behavior and human performance}, 20(2):159--183.

\bibitem[Lichtenstein et~al., 1977]{lichtenstein1977calibration}
Lichtenstein, S., Fischhoff, B., and Phillips, L.~D. (1977).
\newblock {\em Calibration of probabilities: The state of the art}, volume~16
  of {\em Theory and Decision Library}, pages 275--324.
\newblock Springer Netherlands.

\bibitem[Lobo and Yao, 2010]{lobo2010human}
Lobo, M.~S. and Yao, D. (2010).
\newblock Human judgement is heavy tailed: Empirical evidence and implications
  for the aggregation of estimates and forecasts.
\newblock Available at
  \url{http://sousalobo.com/researchfiles/Lobo_Yao_MS_11.pdf}.
\newblock (Working paper).

\bibitem[McCullagh and Nelder, 1989]{mccullagh1989generalized}
McCullagh, P. and Nelder, J.~A. (1989).
\newblock {\em Generalized linear models}, volume~37.
\newblock CRC press, 2nd edition.

\bibitem[Mellers et~al., 2014]{mellers2014psychological}
Mellers, B., Ungar, L., Baron, J., Ramos, J., Gurcay, B., Fincher, K., Scott,
  S.~E., Moore, D., Atanasov, P., Swift, S.~A., Murray, T., Stone, E., and
  Tetlock, P.~E. (2014).
\newblock Psychological strategies for winning a geopolitical forecasting
  tournament.
\newblock {\em Psychological Science}, 25(5):1106--1115.

\bibitem[Moore and Healy, 2008]{moore2008trouble}
Moore, D.~A. and Healy, P.~J. (2008).
\newblock The trouble with overconfidence.
\newblock {\em Psychological Review}, 115(2):502.

\bibitem[Moore and Klein, 2008]{moore2008use}
Moore, D.~A. and Klein, W.~M. (2008).
\newblock Use of absolute and comparative performance feedback in absolute and
  comparative judgments and decisions.
\newblock {\em Organizational Behavior and Human Decision Processes},
  107(1):60--74.

\bibitem[Murphy and Daan, 1984]{murphy1984impacts}
Murphy, A.~H. and Daan, H. (1984).
\newblock Impacts of feedback and experience on the quality of subjective
  probability forecasts. comparison of results from the first and second years
  of the zierikzee experiment.
\newblock {\em Monthly Weather Review}, 112(3):413--423.

\bibitem[Murphy and Winkler, 1977a]{murphy1977can}
Murphy, A.~H. and Winkler, R.~L. (1977a).
\newblock Can weather forecasters formulate reliable probability forecasts of
  precipitation and temperature.
\newblock {\em National Weather Digest}, 2(2):2--9.

\bibitem[Murphy and Winkler, 1977b]{murphy1977reliability}
Murphy, A.~H. and Winkler, R.~L. (1977b).
\newblock Reliability of subjective probability forecasts of precipitation and
  temperature.
\newblock {\em Applied Statistics}, 26(1):41--47.

\bibitem[Murphy and Winkler, 1987]{murphy1987general}
Murphy, A.~H. and Winkler, R.~L. (1987).
\newblock A general framework for forecast verification.
\newblock {\em Monthly Weather Review}, 115(7):1330--1338.

\bibitem[Parunak et~al., 2013]{parunak2013characterizing}
Parunak, H. V.~D., Brueckner, S.~A., Hong, L., Page, S.~E., and Rohwer, R.
  (2013).
\newblock Characterizing and aggregating agent estimates.
\newblock In {\em Proceedings of the 2013 International Conference on
  Autonomous Agents and Multi-agent Systems}, pages 1021--1028, Richland, SC.
  International Foundation for Autonomous Agents and Multiagent Systems.

\bibitem[Raftery et~al., 1997]{raftery1997bayesian}
Raftery, A.~E., Madigan, D., and Hoeting, J.~A. (1997).
\newblock Bayesian model averaging for linear regression models.
\newblock {\em Journal of the American Statistical Association},
  92(437):179--191.

\bibitem[Ranjan and Gneiting, 2010]{Ranjan08}
Ranjan, R. and Gneiting, T. (2010).
\newblock Combining probability forecasts.
\newblock {\em Journal of the Royal Statistical Society: Series B (Statistical
  Methodology)}, 72(1):71--91.

\bibitem[Ravishanker and Dey, 2001]{ravishanker2001first}
Ravishanker, N. and Dey, D.~K. (2001).
\newblock {\em A first course in linear model theory}.
\newblock CRC Press.

\bibitem[Rowse et~al., 1974]{rowse1974comparison}
Rowse, G.~L., Gustafson, D.~H., and Ludke, R.~L. (1974).
\newblock Comparison of rules for aggregating subjective likelihood ratios.
\newblock {\em Organizational Behavior and Human Performance}, 12(2):274--285.

\bibitem[Satop{\"a}{\"a} et~al., 2014a]{satopaa}
Satop{\"a}{\"a}, V.~A., Baron, J., Foster, D.~P., Mellers, B.~A., Tetlock,
  P.~E., and Ungar, L.~H. (2014a).
\newblock Combining multiple probability predictions using a simple logit
  model.
\newblock {\em International Journal of Forecasting}, 30(2):344--356.

\bibitem[Satop{\"a}{\"a} et~al., 2014b]{satopaa2014probability}
Satop{\"a}{\"a}, V.~A., Jensen, S.~T., Mellers, B.~A., Tetlock, P.~E., Ungar,
  L.~H., et~al. (2014b).
\newblock Probability aggregation in time-series: Dynamic hierarchical modeling
  of sparse expert beliefs.
\newblock {\em The Annals of Applied Statistics}, 8(2):1256--1280.

\bibitem[Satop{\"a}{\"a} et~al., 2015]{satopaamodeling}
Satop{\"a}{\"a}, V.~A., Pemantle, R., and Ungar, L.~H. (2015).
\newblock Modeling probability forecasts via information diversity.
\newblock {\em The Journal of the American Statistical Association (Theory \&
  Methods) (arXiv:1406.2148) (In Press)}.

\bibitem[Satop{\"a}{\"a} and Ungar, 2015]{satopaa2015combining}
Satop{\"a}{\"a}, V.~A. and Ungar, L.~H. (2015).
\newblock Combining and extremizing real-valued forecasts.
\newblock {\em arXiv:1506.06405 (Under Review)}.

\bibitem[Savage, 1971]{savage1971elicitation}
Savage, L.~J. (1971).
\newblock Elicitation of personal probabilities and expectations.
\newblock {\em Journal of the American Statistical Association},
  66(336):783--801.

\bibitem[Soll, 1996]{soll1996determinants}
Soll, J.~B. (1996).
\newblock Determinants of overconfidence and miscalibration: The roles of
  random error and ecological structure.
\newblock {\em Organizational Behavior and Human Decision Processes},
  65(2):117--137.

\bibitem[Tanaka and Nakata, 2014]{tanaka2014positive}
Tanaka, M. and Nakata, K. (2014).
\newblock Positive definite matrix approximation with condition number
  constraint.
\newblock {\em Optimization Letters}, 8(3):939--947.

\bibitem[Ungar et~al., 2012]{ungar2012good}
Ungar, L., Mellers, B., Satop{\"a}{\"a}, V., Tetlock, P., and Baron, J. (2012).
\newblock The good judgment project: A large scale test of different methods of
  combining expert predictions.
\newblock The Association for the Advancement of Artificial Intelligence
  Technical Report FS-12-06.

\bibitem[Won and Kim, 2006]{won2006maximum}
Won, J.~H. and Kim, S.-J. (2006).
\newblock Maximum likelihood covariance estimation with a condition number
  constraint.
\newblock In {\em Signals, Systems and Computers, 2006. ACSSC'06. Fortieth
  Asilomar Conference on}, pages 1445--1449. IEEE.

\bibitem[Yates, 1990]{yates1990judgment}
Yates, J.~F. (1990).
\newblock {\em Judgment and decision making}.
\newblock Prentice-Hall, Inc, illustrated edition.

\end{thebibliography}

\end{document}